\documentclass[%
 aip,
amsmath,amssymb,
reprint,%
nofootinbib
]{revtex4-1}
\usepackage[utf8]{inputenc}
\usepackage[T1]{fontenc}
\usepackage{mathptmx}
\usepackage{etoolbox}
\usepackage{lipsum}
\usepackage{graphicx}
\usepackage{amssymb}
\usepackage{url}
\usepackage{soul}
\usepackage[labelformat=simple]{subcaption}

\usepackage{float}
\usepackage{calligra}
\usepackage{calrsfs}
\usepackage{xcolor}
\usepackage{enumitem}
\usepackage{graphicx}
\usepackage{dcolumn}
\usepackage{bm}
\usepackage{tabularx,ragged2e,booktabs,caption}
\usepackage{xcolor}
\usepackage[left=2cm,right=2cm,top=2cm,bottom=2cm]{geometry}

\definecolor{LWcolor}{RGB}{0,0,255} 

\newcommand{\mycomment}[1]{}
\usepackage{icomma}

\setcitestyle{numbers,square}

\definecolor{HGcolor}{RGB}{255,20,147} 

\definecolor{GScolor}{RGB}{0,0,0} 
\newcommand{\gs}[1]{{\color{GScolor}{#1}}}

\makeatletter
\def\@email#1#2{%
 \endgroup
 \patchcmd{\titleblock@produce}
  {\frontmatter@RRAPformat}
  {\frontmatter@RRAPformat{\produce@RRAP{*#1\href{mailto:#2}{#2}}}\frontmatter@RRAPformat}
  {}{}
}%
\makeatother
\begin{document}
\title{Large Eddy Simulations of Fully-Developed Turbulent Flows Over Additively Manufactured Rough Surfaces} 

\author{Himani Garg}
\author{Lei Wang}%
\author{Guillaume Sahut}
\author{Christer Fureby}
\email{himani.garg@energy.lth.se}
 \affiliation{$^1$Lund University$,$ Department of Energy Sciences $,$ P$.$O$.$ Box 118$,$ SE-22100 Lund$,$ Sweden}
\begin{abstract}
\section*{Abstract}
In the last decade, with the growing demand for efficient and more sustainable products that reduce our CO$_2$ footprint, progresses in Additive Manufacturing (AM) have paved the way for optimized heat exchangers, whose disruptive design will heavily depend on predictive numerical simulations. Typical AM rough surfaces show limited resemblance to the artificially constructed rough surfaces that have been the basis of most prior fundamental research on turbulent flow over rough walls. Hence, current wall models used in steady and unsteady three-dimensional (3D) Navier-Stokes simulations do not consider such characteristics. Therefore, a high-fidelity Large Eddy Simulation (LES) database is built to develop and assess novel wall models for AM. This article investigates the flow in rough pipes built from the surfaces created using AM techniques at Siemens based on Nickel Alloy IN939 material. We developed a code to generate the desired rough pipes from scanned planar surfaces. We performed high-fidelity LES of turbulent rough pipe flows at Reynolds number, $Re=11,700$, to reveal the influence of roughness parameters on turbulence, mainly the average roughness height and the Effective Slope. The equivalent sand-grain roughnesses, $k_s$, of the present AM rough surfaces are predicted using the Colebrook correlation. The main contributors to the skin friction coefficient are found to be turbulence and drag forces. In the present study, the existence of a logarithmic layer is marked even for high values of $k_s$. The mean flow, the velocity fluctuations, and the Reynolds stresses show turbulence's strong dependence on the roughness topography. Profiles of turbulence statistics are compared by introducing an effective wall-normal distance defined as zero-plane displacement. The effective distance collapses the shear stresses and the velocity fluctuations outside the roughness sublayer; thus, Townsend's similarity of the streamwise mean velocity is marked for the present roughnesses. Furthermore, a mixed scaling is introduced to improve the collapse of turbulence statistics in the roughness sublayer. In addition, an attempt to investigate surface roughness's impact on flow physics using the acquired LES results based on quadrant analysis of the Reynolds shear stresses and anisotropy of turbulence is made.   
\end{abstract}

\maketitle

\section{Introduction}
With the growing demand for efficient and more sustainable products that reduce our CO$_2$ footprint, there is an urgent need for a new generation of cooling components that can deliver superior thermal performance while
minimizing the cost, size, manufacturing time, and weight. Micro-components and even micro-devices are increasingly used in numerous industrial applications from biomedicine to fuel cells. Most of these devices have microfluidic systems with various functionalities, from process to cooling.  Unfortunately, research on micro- and mini-channel heat transfer have yet to realize its full potential. The market penetration has been limited due to the high-pressure drop associated with narrow flow channels, flow maldistribution issues, and high manufacturing costs.

One way to increase heat transfer performance is to increase the surface area to volume ratio using micro- and mini-channels. Additive Manufacturing (AM) offers possibilities to produce completely novel designs, which are not feasible or economically viable with traditional subtractive manufacturing. 
Metal-based AM has been a promising technology in the design of advanced cooling strategies for gas turbines. For internal cooling, the goal is to maximize the heat transfer coefficient while minimizing the pressure drop to ensure cooling air is used most efficiently. However, a major challenge arising when using AM for producing metal parts is the large surface roughness resulting from the layer-by-layer buildup process. The effects of surface roughness in the cooling channels are twofold. Firstly, it augments the convective heat transfer by disturbing the boundary layer. Secondly, it produces a significant increase in pressure loss, which may reduce the cooling air's mass flow rate. More importantly, the high-pressure burnt gas may rush into the cooling passage and destroy the hot components. Therefore, it is critical to understand and accurately predict the roughness effects on the flow and heat transfer in additively manufactured channels, which constitutes the main motivation behind this work. 

One of the fundamental questions is how to quantify the effects of surface roughness on the mean velocity profiles. \citet{hama1954boundary} and \citet{clauser1954} showed that
wall roughness usually gives rise to significant form drag and skin friction, increasing turbulence and leading to a downward shift of the mean velocity profile, commonly known as the roughness function. Almost a century ago, the most famous investigation regarding the effect of roughness on the skin friction coefficient in rough-wall turbulence was carried out by \citet{nikuradse1950laws}. In his seminal paper, a large number of measurements of a pressure drop in pipes roughened by homogeneous sand grains revealed that if the Reynolds number is sufficiently high, the skin friction coefficient only depends on an equivalent sand-grain roughness height, $k_s$. 
His work was further extended by \citet{colebrook1939correspondence} and \citet{moody1944friction}, related the pressure drop in a pipe to the relative roughness (ratio of roughness height to pipe diameter) and Reynolds number. 
Results by \citet{allen2007turbulent} dictate that the pressure drop in the Moody chart is overestimated in transitionally rough regimes for honed and commercial steel pipes. This clearly suggests that the Colebrook roughness function used in the formulation of the Moody chart may not be relevant to a wide range of roughness of engineering interest, according to \citet{flack2010review}.
Nevertheless, the Moody chart is a powerful tool for assessing the wall roughness effect on pressure loss as long as the equivalent roughness to the actual roughness is known a priori. However, in terms of additively manufactured roughness, the equivalent sand-grain roughness height is usually unknown. Consequently, estimating equivalent roughness to predict skin friction is a daunting task, and many studies have dedicated their efforts to exploring the correlation to predict the equivalent sand-grain roughness height based on the existing roughness topology  \citep{schlichting1961boundary,dvorak1969calculation,coleman1984,van2002analysis}.    
For the prediction of the equivalent sand-grain roughness height, \citet{Dirling1973} and later on, \citet{van2002analysis} proposed a roughness parameter including a roughness density and a shape parameter. 
They reported that the suggested correlation could also apply to walls of non-uniform, three-dimensional roughness with irregular geometries. Furthermore, this correlation was reported to provide an adequate estimate of real turbine blade roughness by \citet{bons2002st}.
Moreover, based on statistical measures of surface elevation, several kinds of correlations for the equivalent sand-grain roughness height have been proposed \citep{musker1980universal,flack2010review,flack2016skin}. 

\citet{townsend1980structure} wall similarity hypothesis states that, at a sufficiently high Reynolds number, turbulent structures located above the roughness sublayer are independent of wall roughness and viscosity. The roughness sublayer, i.e., the region located directly above the roughness and extending approximately to $5k_s$ from the wall, in which the roughness length scales directly impact the turbulent motions, has been a subject of in-depth interest. Experimental investigations by \citet{Raupach1991} and \citet{flack2005experimental} demonstrated strong experimental support for outer-layer similarity in the turbulence structure over smooth and rough walls with regular roughnesses. 
\citet{jimenez2004turbulent} stated that the validity of outer-layer similarity depends on the relative roughness of the flow, $k/\delta$, where $k$ is the roughness height and $\delta$ is the boundary layer thickness. If the roughness height is small compared to the boundary layer thickness, $k/\delta < 0.025$, the effect of the roughness is confined to the inner layer, and the wall similarity hypothesis will hold. If, on the other hand, the roughness height is large compared to the boundary layer thickness, $k/\delta > 0.025$, roughness effects on the turbulence may extend across the entire boundary layer, and the concept of wall similarity will be invalid. 

On the other hand, numerical simulations have been recognized as a powerful tool for detailing fundamental flow physics over rough surfaces with different topologies. 
Both Direct Numerical Simulation (DNS) and Large Eddy Simulation (LES) for flows over spanwise extended transverse ribs \citep{miyake2001direct,leonardi2003direct,ashrafian2006structure,
ikeda2007direct,jin2015structure} and three-dimensional roughness with regular arrangement \citep{bhaganagar2004effect,orlandi2008direct,chatzikyriakou2015dns}, such as cubes, cylinders, transverse wedges, and longitudinal wedges with the staggered arrangement, have been performed. The progress in computer technology also enabled researchers to perform DNS of turbulence over rough surfaces with complex geometries, including irregular and random surfaces \citep{napoli2008effect,de2010turbulence,yuan2014estimation,bhaganagar2015characterizing,Forooghi2017,forooghi2018direct} and real, scanned surfaces \citep{busse2015direct,busse2017reynolds,forooghi2018dns}.
In particular, \citet{napoli2008effect} considered two-dimensional irregular corrugated walls to discuss the influence of surface slope on turbulence. In their study, the roughness function for the rough surfaces with an Effective Slope ($ES$) lower than a certain threshold increased with $ES$. However, the dependence on $ES$ became weaker for rough surfaces with $ES$ higher than the threshold. A few systematic experimental \citep{schultz2009turbulent} and numerical \citep{2005chan2015systematic,Forooghi2017} investigations further confirmed the dependence of turbulence on $ES$. 
In addition, \citet{bhaganagar2015characterizing} found from their DNS study on flow over three-dimensional irregular rough surfaces that the skewness and the kurtosis could be effective for characterizing the roughness function. In their study, the spacing between the roughness elements was not sensitive to the roughness function for irregular rough walls. The importance of the skewness has also been reported by DNS studies of surfaces with randomly distributed semi-ellipsoid/cone roughness \citep{Forooghi2017} and realistic rough surfaces \citep{yuan2014estimation}.

A great deal of effort has been made to examine the relation between roughness topological parameters and an increase in skin friction. However, the surface roughness produced by the AM technique is characterized by spatial non-uniformity, which is affected by a variety of parameters such as profile curvature, layer thickness, laser power, sample orientation, metallic composition, and particle size. Therefore, AM roughness is considered to be quite different from regular, random, and artificial roughnesses. To date, very few experimental and numerical studies have been carried out to investigate the convective heat transfer properties in mini-channels made with AM \citep{stimpson2016roughness,stimpson2017scaling,2019Synder,snyder2020tailoring,mcclain2021flow,FAVERO2022106128}. \citet{stimpson2016roughness,stimpson2017scaling} focused on characterizing the morphology and quantifying the roughness of micro-channel surfaces made with Direct Metal Laser Sintering (DMLS) to develop correlations that relate the physical roughness specifications to the effect of the roughness on the flow friction and heat transfer. 
In addition, \citet{snyder2020tailoring} successfully demonstrated the concept of tailoring surface roughness using AM process parameters to improve the performance of a generic micro-channel cooling design. Very recently, the first DNS study of flows in AM-made channels has been published by \citet{altland_zhu_mcclain_kunz_yang_2022}. However, their focus was more placed on super-rough channels where large-scale roughness size is comparable to the channel height. The flow in such super-rough channels is more strongly influenced by the roughness on both surfaces and therefore lacks streamwise and spanwise homogeneity everywhere in the channel. 
This study is very interesting but does not align with the aim of the present study, where we focus on AM-made surfaces of generic nature with small-scale roughness.   

Despite these dedicated initial efforts to understand AM roughness, all these studies mainly focus on heat transfer and pressure drop measurement, documenting significant augmentation of skin friction factor and Nusselt number. 
How AM roughness impacts the internal flow behavior compared to artificial, random, and regular roughnesses is still an open question. In most of these studies, the provided correlations compare average roughness height, $R_a$, with $k_s$, which is optimal for some engineering applications but can not be generalized. 
This choice could be because, in experiments, the measurement of $R_a$ is straightforward, whereas the measurement of higher-order statistics is quite cumbersome. 
Furthermore, based on previous literature, one expects higher-order moments to correlate better with $k_s$ than $R_a$. Therefore, how well AM roughness scales with other statistical measurements of roughness, such as standard deviation ($R_q$), skewness ($s_k$), kurtosis ($k_u$), and $ES$ remains to be understood.
Overall, there is a large gap in understanding the dynamical impact of AM roughness on turbulent flow structures, such as mean flow, turbulent fluctuations, shear stresses, and secondary flow motions. 
We need to find out whether the logarithmic layer survives and, even if so, how well it scales with equivalent sand-grain roughness. Finally, the performance of well-established LES subgrid-scale models for these types of roughnesses is unknown. To answer these research questions and fill the gap in our understanding, we conduct Wall-Resolved LES (WRLES) in additively manufactured rough channels.
A systematic study of the relation between the roughness characteristics and the equivalent roughness is carried out to reveal how roughness characteristics affect the equivalent roughness. 
As of the writing of this article (to the best of the authors' knowledge), this is the first time LES has been used to simulate a turbulent additively-manufactured rough wall pipe flow with grid-conforming three-dimensional roughness elements.

\section{LES methodology}
In LES, an attempt is made to capture the large-scale unsteady motions which carry the bulk of the mass, momentum, and energy in a flow. The aim is to approximately resolve the Taylor scales, the length scales on which these \emph{resolved-scale} motions occur depending primarily on the local grid resolution. Small-scale motions that occur on length scales smaller than a given filtering length, typically taken as the local grid spacing, cannot be captured, and their effects on the resolved-scale motions must be modeled.
The filtered incompressible Navier-Stokes Equations (NSE) are given by
\begin{align}
\dfrac{\partial\overline{u}_i}{\partial x_i} &= 0, \label{eq:2}\\
\dfrac{\partial\overline{u}_i}{\partial t} + \dfrac{\partial}{\partial x_j}(\overline{u}_i\overline{u}_j) &= -\dfrac{1}{\rho}\dfrac{\partial\overline{p}}{\partial x_i}-\dfrac{\partial\tau^{\text{sgs}}_{ij}}{\partial x_j}+\frac{\partial}{\partial x_j}\left[\nu \left(\frac{\partial \overline{u}_i}{\partial x_j}+\frac{\partial \overline{u}_j}{\partial x_i}\right) \right], \label{eq:2bis}
\end{align}%
where $\overline{{u}}_{i}$ is the filtered velocity, $\overline{p}$, the filtered pressure, $\nu$, the kinematic viscosity, and $\tau^{\text{sgs}}_{ij} = \overline{u_i u_j}-\overline{u_i}\,\overline{u_j}$, the subgrid-scale stress tensor. 

In WRLES approach used in this study, a subgrid model is used for the subgrid stress tensor combined with a specific model to represent the proximity to the wall \cite{boussinesq1897}.
The most frequently used subgrid models are eddy-viscosity-based models. 
For the present study, we adopted the Wall-Adapting Local Eddy viscosity (WALE) model \cite{nicoud1999subgrid}. 
In WRLES the aim is to resolve the large-scale, energy-containing flow structures in the outer boundary layer while modeling the flow structures in the inner layer. To this end, an algebraic subgrid wall model based on Spalding’s law-of-the-wall \cite{Spaling1961} is used to compute the subgrid viscosity, $\nu_{\text{sgs}}$, in the inner wall layer to provide a smooth, consistent flow profile adjacent to the wall.

\section{Numerical methods}

OpenFOAM version 7 (OF7) is used to simulate the turbulent pipe flow with AM rough walls. The NSE are solved using the finite volume method with a second-order cell-centered discretization scheme for convective and diffusive fluxes. Time stepping is performed using the second-order accurate implicit Adams-Bashforth method \cite{butcher2016numerical} with a maximum Courant–Friedrichs–Lewy (CFL) number of 1 for numerical stability. The pimpleFoam solver is used in which the pressure-momentum system is decoupled using the Pressure Implicit with Splitting of Operator (PISO) algorithm for LES \cite{weller1998tensorial}. The resulting pressure equation is solved using a Generalized Geometric-Algebraic Multigrid (GAMG) method \cite{GAMG} with a Diagonal Incomplete Cholesky (DIC) smoother \cite{van1996matrix}. Finally, velocity is solved using a Preconditioned BiConjugate Gradient (PBiCG) solver \cite{press2007numerical} with a Diagonal Incomplete Lower-Upper (DILU) preconditioning method \cite{van1996matrix}. 
\section{Simulation setup}

\subsection{Geometry creation: development of a dedicated open-source code}
\label{STL Code}

\begin{figure*}
     \centering
     \begin{subfigure}[b]{0.45\textwidth}
         \centering
         \includegraphics[width=\textwidth]{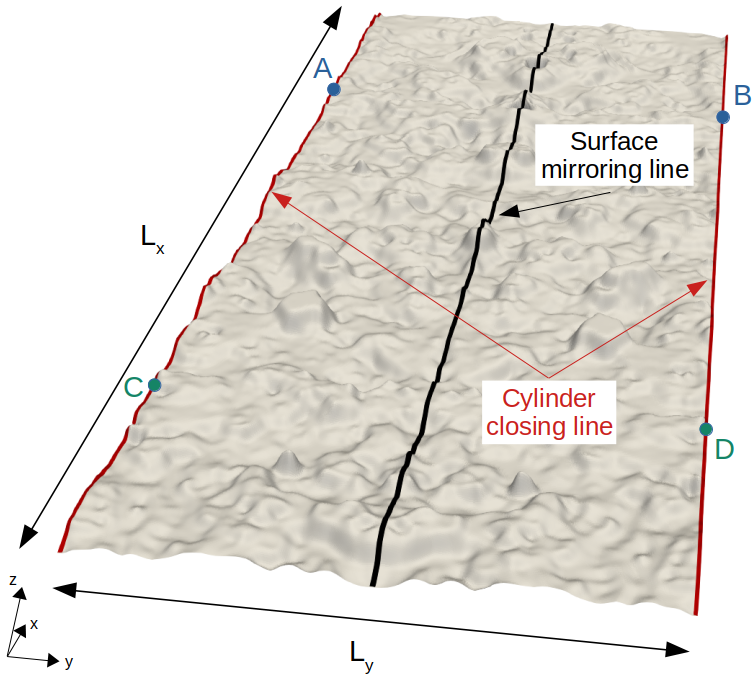}
         \caption{}
         \label{fig:code_plane}
     \end{subfigure}
     \begin{subfigure}[b]{0.45\textwidth}
         \centering
         \includegraphics[height=6.8cm]{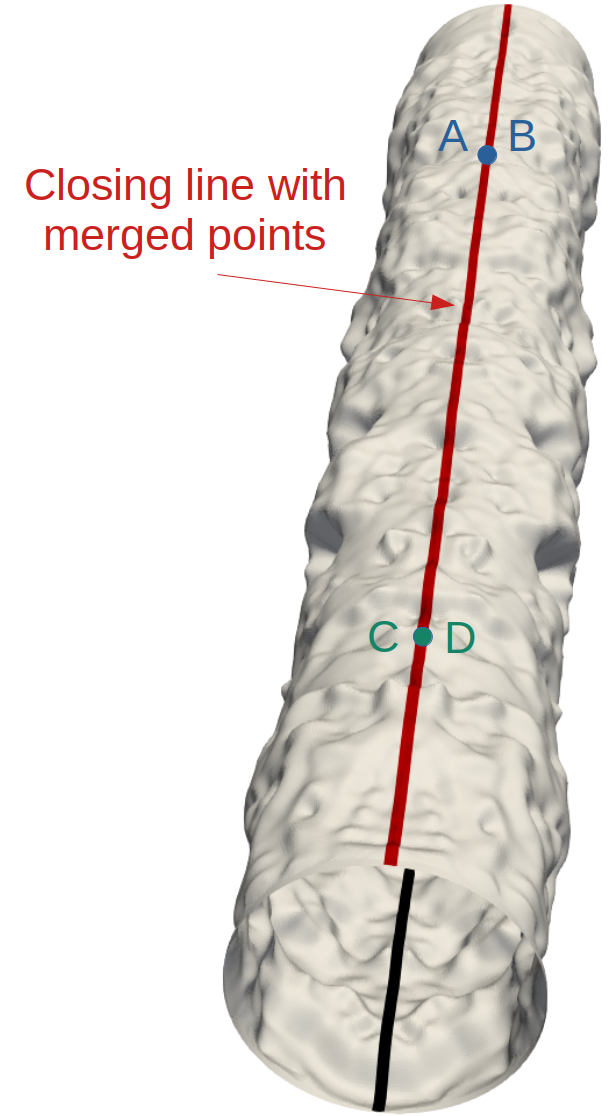}
         \caption{}
         \label{fig:code_cylinder}
     \end{subfigure}
     \caption{Sketch of \subref{fig:code_plane}, a planar rough surface mirrored along the surface mirroring line and \subref{fig:code_cylinder}, the same surface wrapped around the direction of the surface mirroring line where superimposed points of the cylinder closing line are merged, e.g., points $A$ and $B$, and points $C$ and $D$. Prior to bending (point (2)), the two smallest sides of the surface have been flattened using Blender to ease the use of periodic boundary conditions.}
     \label{fig:code_plane2cylinder}
\end{figure*}

The AM rough surfaces used in this study are made of Nickel Alloy IN939 and provided by Siemens. They have been obtained from planar rough surfaces scanned with a light microscope 
and written in STereoLithography (STL) file format. Figure \ref{fig:code_plane2cylinder} shows the transformation of a rough plane into a rough pipe. This operation consists of three steps: (1) mirroring the surface along the rotation axis ($x$-axis in Fig. \ref{fig:code_plane2cylinder}) so that the two surface borders can be merged easily after rotation (see cylinder closing line in Fig. \ref{fig:code_plane}); 
(2), applying a rotation to all points of the STL file around the rotation axis; and (3), ensuring that the cylinder is properly closed along its main axis
, as shown in Fig. \ref{fig:code_cylinder}. 
Since point (1) is relatively standard, we will now focus on points (2) and (3).
Two options have been considered for the creation of rough cylinders. First, one can easily bend a planar surface to produce a cylinder using commercial or open-source computer-aided design software. 
However, 
point (3) may be quite time-consuming. Indeed, the operation of merging points is typically achieved using an absolute threshold distance value such that two points whose separating distance is lower than the threshold value are considered to be merged. The immediate problem here is the difficulty, or even sometimes the impossibility, of determining an appropriate threshold value.
To properly accomplish this operation, one then needs to make many attempts in a trial and error approach and carefully inspect the whole closing line to ensure both that no hole remains and that not too many points have been merged. Another concern about time consumption emerges when this procedure has to be performed a significant number of times, e.g., to gather statistics. Such repetitive actions then become tedious.

To reduce time consumption in the generation of these rough cylinders, we have developed a specific open-source code written in Fortran and available on GitHub\footnote{\url{https://github.com/CoffeeDynamics/STLRoughPipes}}. This code is run in a Unix command-line interpreter and currently takes as arguments the path to the planar rough STL file, a tolerance factor to merge the points along the closing line of the cylinder, and an optional roughness factor to rescale the pipe surface roughness. For all surfaces used in this study, a tolerance factor of 
$0.002$ led to properly closed cylinders. This tolerance factor is inversely proportional to the surface resolution and hence requires particular attention. The ability to specify it in the command line constitutes a valuable ingredient of this code by speeding up the process of building rough cylinders from rough planar surfaces. Furthermore, the possibility to rescale the pipe surface roughness from the command line enables fast parametric studies of the influence of the roughness height for a given surface topography. The code produces a cylinder with a rough surface in STL format, as shown in Fig. \ref{fig:code_cylinder}, as well as some statistical quantities (e.g., average surface roughness, arithmetic mean deviation, skewness, and kurtosis, as will be detailed in Table \ref{table: Roughness parameters intro}, Section \ref{sec:3}). It can then be invoked in a loop over STL files to process a large number of surfaces at once.

\subsection{Initial and boundary conditions, dimensionless numbers}
The AM-roughened pipe flow is simulated at a fixed $Re_b=11,700$, where $Re_b=U_bD/\nu$ is the bulk Reynolds number of the flow, with $U_b$ denoting the bulk velocity, $D$, the diameter of the circumscribed cylinder, and $\nu$, the kinematic viscosity. The flow is assumed to have constant material properties. In addition to $U_b$, $u_\tau$ is introduced in connection to wall-bounded flows, calculated a posteriori based on the pressure gradient, where $u_\tau =\sqrt{0.25D\Delta P/(\rho L_x)}$ is the friction velocity scale and $\Delta P/L_x$ is the pressure gradient. 
Based on the computed value of $u_\tau$, the turbulent friction Reynolds number, $Re_\tau =u_\tau D/(2\nu)$ is defined, yielding the four values $Re_\tau = 448$, $524$, $615$, and $875$ used in this study. In addition, the roughness Reynolds number, $k_s^+$, based on $k_s$, and defined as $k_s^+ = {k_s u_\tau}/{\nu}$ is also considered, resulting in $k_s^+ = 17, 44, 83$, and $422$. This will be visited later. Note that for roughened-wall flows, $u_\tau$ and the skin friction coefficient, $f_d$, are no longer composed solely of the skin-friction drag but actually reflect the total wall drag which is composed of pressure and viscous drag components. 
Here, $ L_x$ denotes the length of the computational domain. The axial, radial, and azimuthal directions are represented by $x$, $r$, and $\theta$, and the corresponding velocity components, by $u_x$, $u_r$, and $u_\theta$. The effective wall-normal distance, i.e., along the radial direction, is denoted by $r = r^0-d$, where $r^0$ is the pipe radius and $d$ is the zero-plane displacement, this will be visited later in Sec. \ref{Sec: Mean flow profiles}. Here, $u_x'$, $u_r'$, and $u_\theta'$ are the velocity fluctuations in the streamwise, radial, and azimuthal directions, respectively. Since the fully-developed turbulent pipe flow with rough walls considered here is homogeneous in the streamwise direction, periodic boundary conditions are also imposed in this direction. No-slip boundary conditions are imposed at the pipe wall for all velocity components, whereas zero-Neumann boundary condition is used for the pressure. An additional force term is introduced into the NSE to mimic the effect of the pressure gradient in the context of periodic boundary conditions in the streamwise direction. The magnitude of this force is determined by the bulk velocity. 
For the present study, we chose the pipe length to equal roughly eight times the pipe diameter, i.e., $L_x=8D$, a sufficiently large ratio to ensure that all statistics considered here are free from periodicity effects. In addition, a mesh of roughly $2$ million cells has been used for all cases, and this choice is motivated by the detailed mesh sensitivity study performed at $Re_b=11,700$ for the smooth pipe flows in \cite{Himani2022}. 
\section{\label{sec:3}Results and discussion}
\subsection{Surface roughness parameters and characterization} \label{sec:SRPC}
\begin{figure*}
\begin{center}
\subfloat[]{
\includegraphics[width=0.5\linewidth]{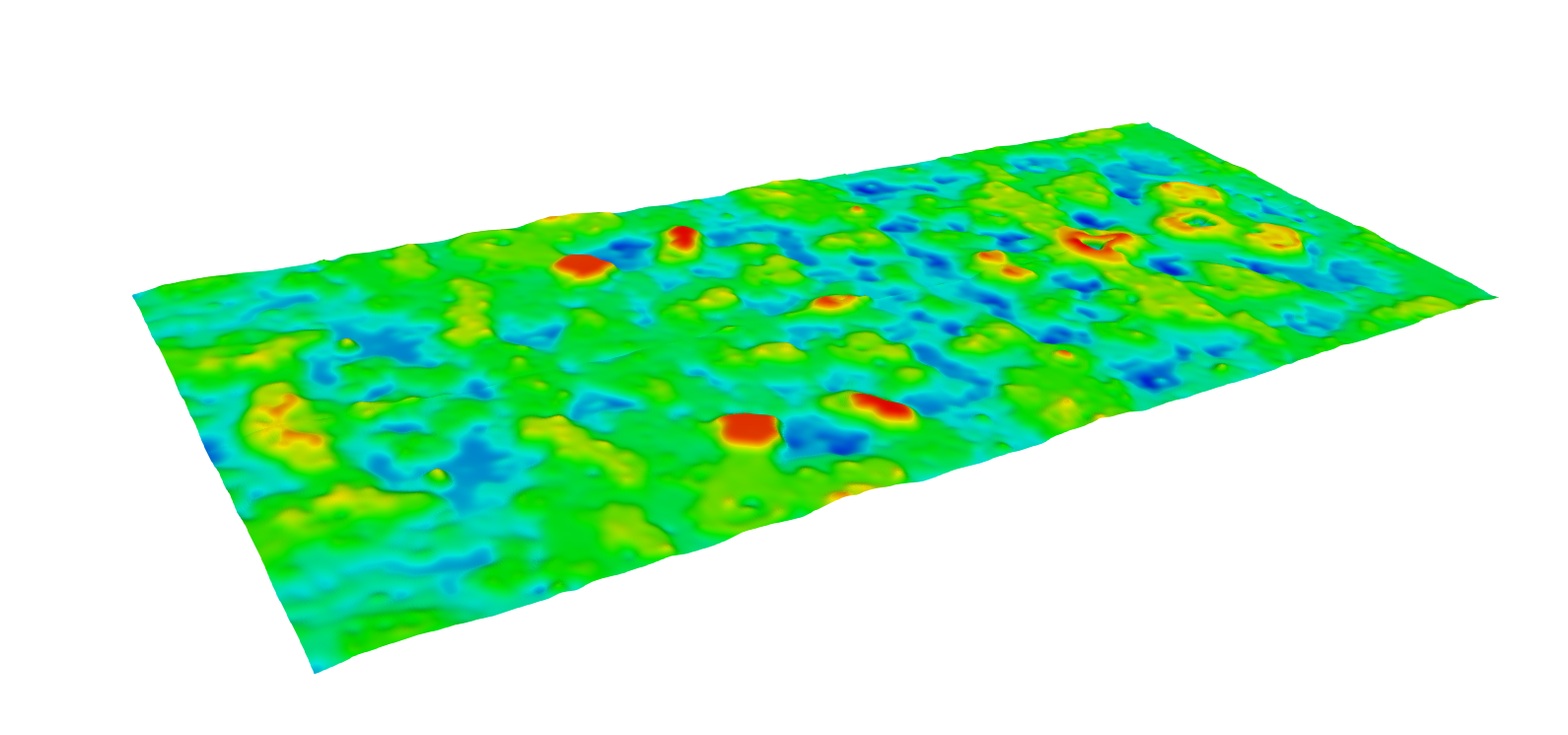}}
\subfloat[]{
\includegraphics[width=0.5\linewidth]{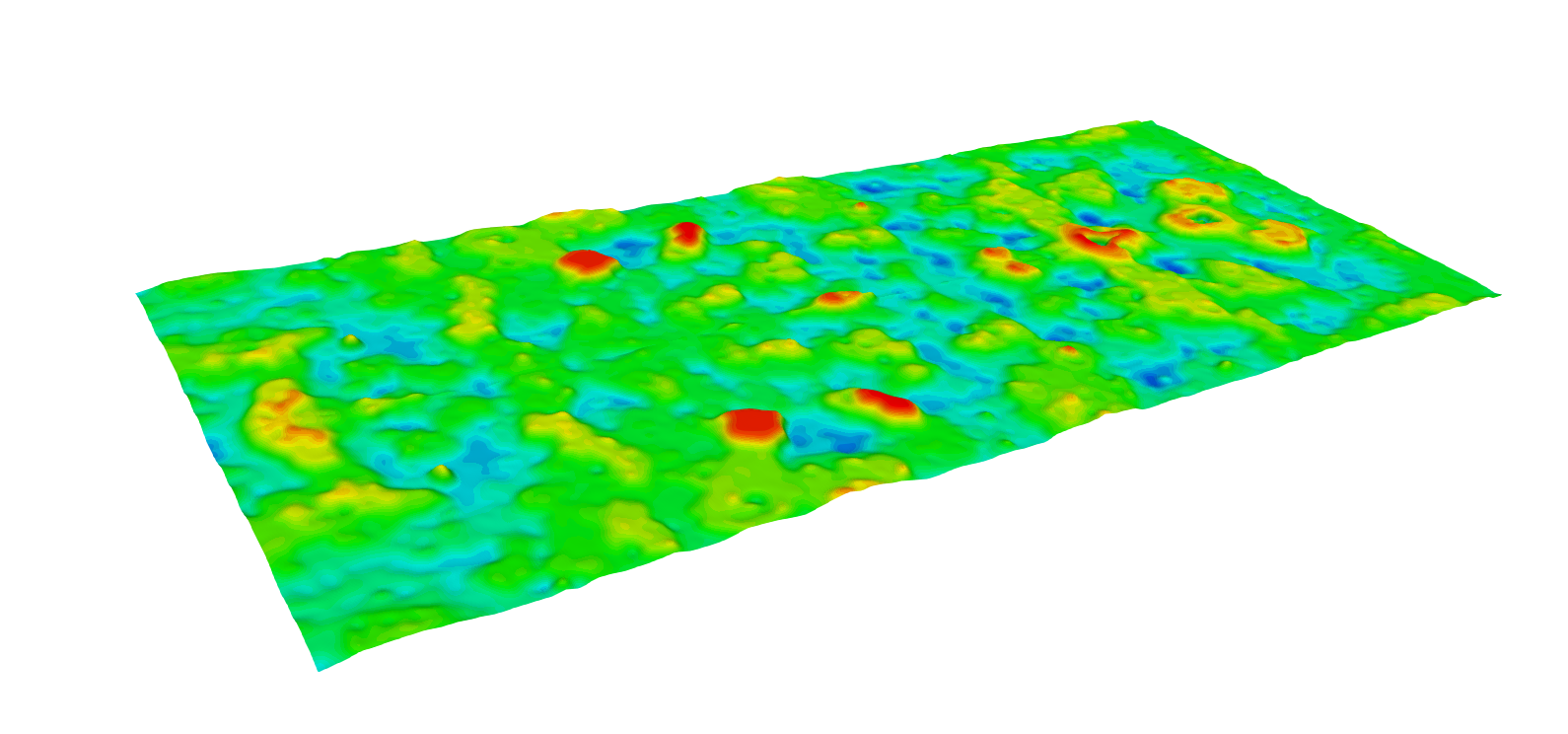}}

\subfloat[]{
\includegraphics[width=0.5\linewidth]{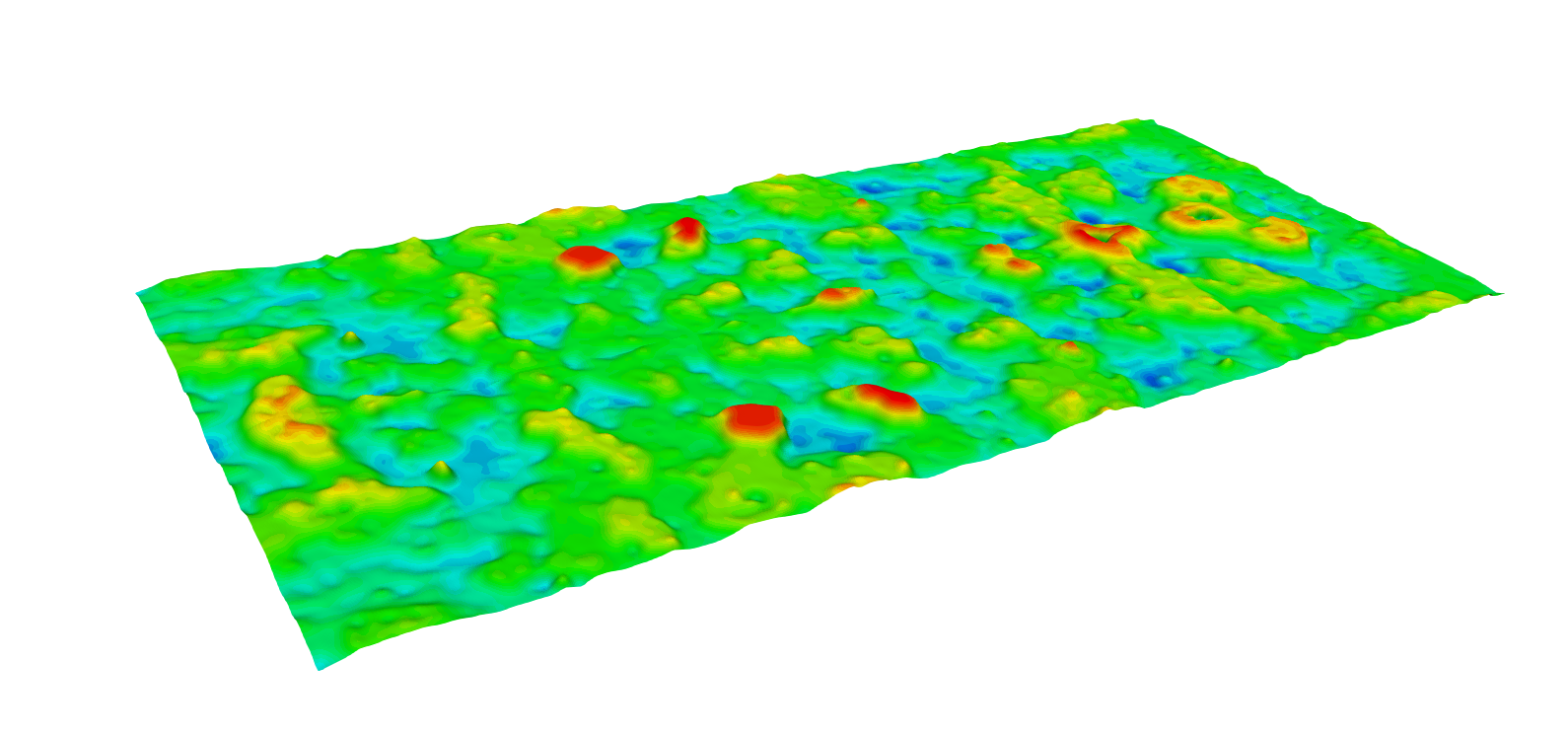}}
\subfloat[]{
\includegraphics[width=0.5\linewidth]{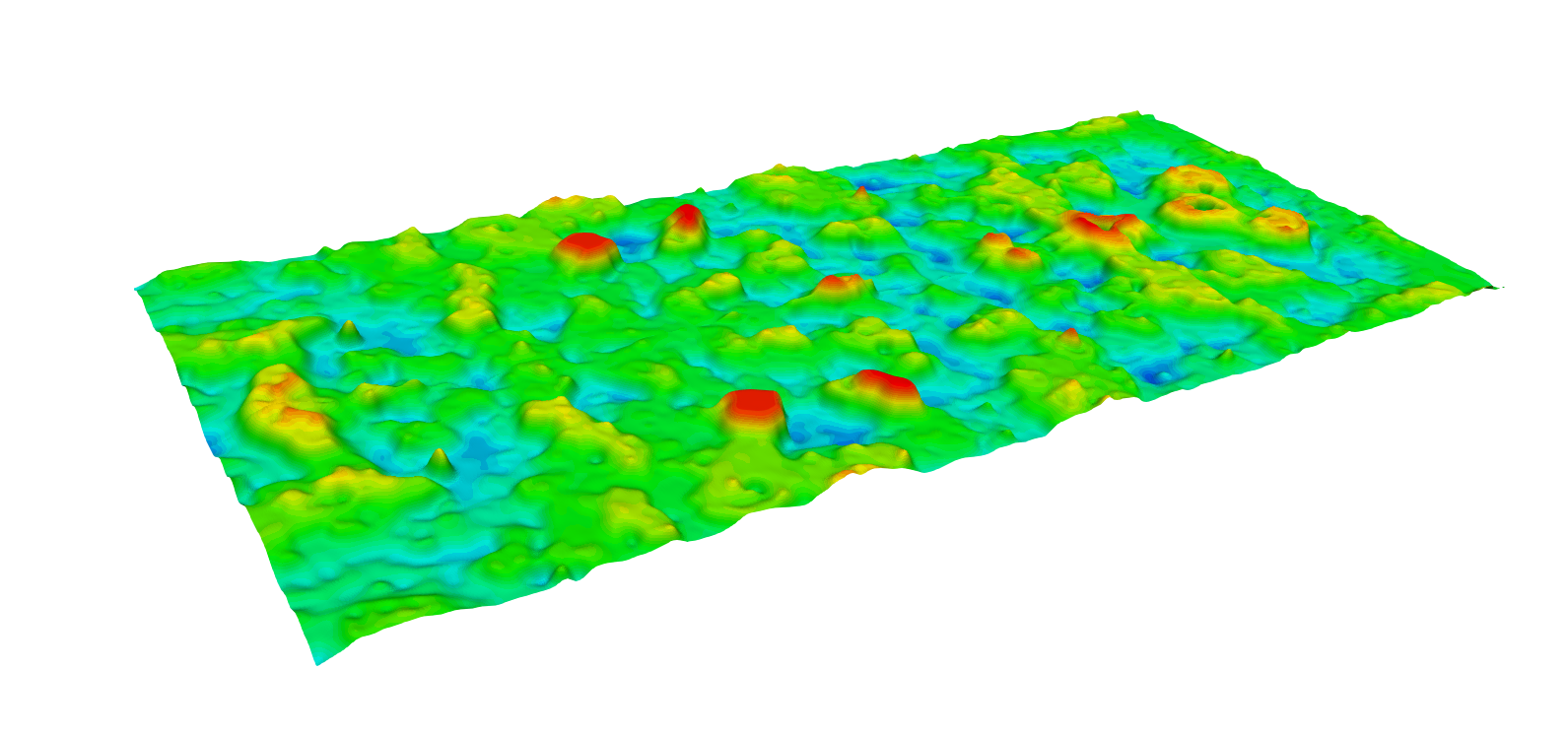}}
\caption{Visualization of AM rough surface height map, $y(x,z)$. The 3D scanned surfaces are extracted from the measurement of 3D printed microchannels at Siemens. Note that the actual scanned surfaces were larger, $2 \times 2$ mm. Due to the unavailability of high resolution, only a subpart of the existing surfaces is considered in this study, roughly $2 \times 0.5$ mm (before mirroring). While moving from (a) to (d), the sampled surface is roughly the same, but the roughness height increases.} 
\label{fig: Rough Surface Heigh Map}
\end{center}
\end{figure*}
Generally, the surface morphology comprises a series of geometrical irregularities randomly span over a smooth surface. Surface roughness characterization is essential in studying many fundamental problems, such as friction and heat transfer, from an engineering point of view. 
The real surface geometry is so complex that a finite number of parameters cannot describe it completely. 
For the present study, we consider four additively manufactured rough surfaces that share roughly the same topography but show different roughness heights, see Fig. \ref{fig: Rough Surface Heigh Map}. We adopted numerous statistical surface parameters to characterize surface roughness \citep{davim2010surface, leach2013characterisation}. Roughness parameters can be calculated either on the profile of a sampling line (two-dimensional (2D) forms) or a sampling area (3D forms). 
We consider the whole surface in the present study, and the 3D roughness parameters are then calculated. The different parameters adopted are summarized in Table \ref{table: Roughness parameters intro}. We also adopt one of the extensive parameters called Effective Slope ($ES$) \citep{napoli2008effect} to distinguish between roughness and waviness surface types, given by
\begin{equation}
 ES =   \frac{1}{L_xL_z} \int_0^{L_z} \int_0^{L_x}\left | \frac{\partial y(x,z)}{\partial x} \right |dx dz,
\end{equation}
where $y(x,z)$ is the roughness amplitude, also known as the height function, while $L_x$ and $L_z$ are the sampling lengths in the streamwise and spanwise directions, respectively. $ES$ is defined as the average value of the slope magnitude of the roughness. 

Roughness parameters for the cases presented here and obtained with the newly developed code mentioned in Sec. \ref{STL Code} are reported in Table \ref{table: Roughness parameters values}. For the current study, the values of $s_k$ and $k_u$ are fixed for all rough surfaces. The critical value of $ES$ for considered rough surfaces is mostly found to be much smaller than the critical value of $ES=0.35$ \citep{yuan2014estimation}. \citet{Forooghi2017} showed that, for lower values of $ES$, the surface slope becomes the main controlling factor and suppresses the impact of the topographical parameters listed in Table \ref{table: Roughness parameters intro}. 
Once the simulations reached convergence, we extracted the values of $\Delta P/L_x$, $u_\tau$, $f_d$, and eventually $k_s$ for all roughnesses. These values are documented in Table \ref{table: Different roughness heights}.
 \begin{table*}
 \setlength{\tabcolsep}{1em}
\begin{center}
\caption{Summary of roughness parameters based on roughness height, $y$, with continuous and discrete formulations, where $L_x$ is the surface length, $n$, the number of samples, and $p$, the probability density function.}
\begin{tabular}{l p{60mm} l l l}
\hline \\[0cm]
Parameter & Description & Continuous formulation & Discrete formulation  \\[0.25cm]\hline \\[0.cm]
$R_a$   & Arithmetic average height  & $\frac{1}{L_x}\int_0^{L_x} y(x) dx$ & $\frac{1}{n}\sum_{i=1}^{n} y_i $    \\[0.5cm]
$R_q$  & Root-mean-square roughness height  & $\sqrt{\frac{1}{L_x}\int_0^{L_x} \left \{ y(x)-R_a \right \}^2 dx}$  &    $\sqrt{\frac{1}{n}  \sum_{i=1}^{n}\left\{y_i-R_a\right\}^2}$      \\[0.5cm]
$R_v$   & Maximum depth of valleys below the mean line, within the sampling surface  & $\left |\min y_i\right |+R_a$  & $\left |\min y_i\right |+R_a$      \\[0.5cm]
$R_p$   & Maximum height of peaks above the mean line, within the sampling surface  & $\max y_i - R_a$ & $\max y_i - R_a$       \\ [0.5cm]
$R_{\text{max}}$   &  Maximum height between the highest peak and the deepest valley of the profile  & $R_v+R_p$  & $R_v+R_p$      \\[0.7cm] 
$s_k$   & Skewness   & $\frac{1}{R_q^3} \int_{-\infty}^{+\infty} \left\{y-R_a\right\}^3p(y)dy$   &  $\frac{1}{nR_q^3}\sum_{i=1}^{n}\left\{y_i-R_a\right\}^3$      \\[0.5cm] 
$k_u$   & Kurtosis & $\frac{1}{R_q^4} \int_{-\infty}^{+\infty}\left\{y-R_a\right\}^4p(y)dy$  & $\frac{1}{nR_q^4}\sum_{i=1}^{n}\left\{y_i-R_a\right\}^4$     \\[0.5cm] 
${ES}$   & Effective Slope based on height map $y(x,z)$   & $\frac{1}{L_xL_z} \int_0^{L_z} \int_0^{L_x}\left | \frac{\partial y(x,z)}{\partial x} \right |dx dz$  & Centered difference scheme      \\[0.5cm] \hline
\end{tabular}
\label{table: Roughness parameters intro}
\end{center}
\end{table*}
 \begin{table*}
 \setlength{\tabcolsep}{10pt}
\begin{center}
\caption{Statistical quantities for the four rough surfaces considered. The descriptions and analytical expressions are given in Table \ref{table: Roughness parameters intro}.}
\begin{tabular}{c c c c c c c c c }
\hline \\[0cm] 
Case  & $R_a$ [m] & $R_q$ [m] & $R_v$ [m] & $R_p$ [m] & $R_{\text{max}}$ [m] & $s_k$ [-] & $k_u$ [-] & ${ES}$ [-] \\[0.2cm] \hline \\[0.cm]
C1    & 1.594e-3 &2.819e-3 & 1.217e-2 & 1.418e-2 & 2.636e-2 & 0.599 & 6.107 & 0.102 \\[0.2cm]
C2    & 1.992e-3 & 3.524e-3 & 1.522e-2 & 1.773e-2 & 3.295e-2 & 0.599 & 6.107 & 0.124 \\[0.2cm]
C3    & 2.630e-3 & 4.652e-3 & 2.009e-2 & 2.340e-2 & 4.349e-2 & 0.599 & 6.107 & 0.157 \\[0.2cm]
C4    & 3.984e-3 & 7.048e-3 & 3.043e-2 & 3.546e-2 & 6.589e-2 & 0.599 & 6.107 & 0.216  \\[0.2cm]
\hline
\end{tabular}
\label{table: Roughness parameters values}
\end{center}
\end{table*}

\begin{table*}
\setlength{\tabcolsep}{8pt}
\begin{center}
\caption{Estimation of pressure gradient, $\Delta P/L_x$, friction velocity, $u_\tau$, skin friction factor, $f_{d}$, equivalent sand-grain roughness height, $k_s$, equivalent sand-grain roughness height normalized by $R_a$ and $R_q$, and equivalent sand-grain roughness height, $k_s^{ES}$, employing a correlation based on $ES$, from LES simulations. The value of the pressure gradient is extracted at the end of each iteration and averaged over long time series.}
\begin{tabular}{c c c c c c c c}
\hline \\[0cm] 
Case  & $\Delta P/L_x$ [Pa.m$^{-1}$]  & $u_ \tau$ [m/s]  & $f_{d}$ [-]  & $k_s$ [m]& $k_s/R_a$ [-] & $k_s/R_q$ [-] & $k_s^{ES}$ [m]\\[0.2cm] \hline \\[0.cm]
C1   & 0.068 & 0.067 & 0.051   & 0.005 & 3.201 & 1.806  & 0.004\\[0.2cm]
C2   & 0.087 & 0.077 & 0.069   & 0.011 & 5.760 & 3.253  & 0.006\\[0.2cm]
C3   & 0.106 & 0.087 & 0.087   & 0.019 & 7.262 & 4.106 & 0.008\\[0.2cm]
C4   & 0.194 & 0.119 & 0.180   & 0.071 & 17.874 & 10.103  & 0.014\\[0.2cm]
\hline
\end{tabular}
\label{table: Different roughness heights}
\end{center}
\end{table*}


\subsection{Mean flow profiles}
\label{Sec: Mean flow profiles}
\begin{figure*}
\begin{center}
\includegraphics[width=1\linewidth]{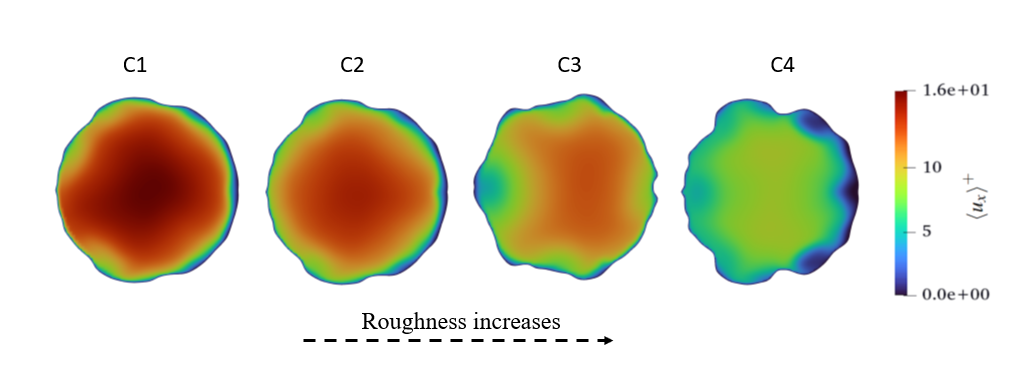}
\caption{Cross-sectional view of the mean streamwise velocity, normalized by the friction velocity, $u_\tau$, for different rough surfaces: (\textit{a}) C1, (\textit{b}) C2, (\textit{c}) C3, and (\textit{d}) C4.} 
\label{fig: flow visualizations}
\end{center}
\end{figure*}

All the statistics obtained in this study have been averaged in time and along the streamwise and azimuthal directions. Moreover, due to the symmetry between the pipe wall's lower and upper parts, the statistics are symmetrically averaged with respect to the mid-plane of the pipe.

One of the most important effects of the wall roughness on the turbulent flow is the reduction of the streamwise velocity with respect to smooth-wall conditions, with the consequent modification of the friction coefficient. In the log region, at a sufficient distance from the roughness elements, the stream-wise velocity profile can be expressed as
\begin{equation}
\left<u_x\right>^+ = \frac{1}{k} \text{log}(y^+)+C_s-\Delta U^+,
\label{eq:res 1}
\end{equation}
 where $\left<u_x\right>^+$ is the non-dimensional time-averaged streamwise velocity, $k$, the von Kármán constant, $y^+$, the non-dimensional wall-normal distance, $C_s$, a constant equal to $5.2-5.5$ for channel flows depending on the Reynolds number, and $\Delta U^+$, the roughness function. The first two terms of the right-hand side come from the expression valid for a smooth wall. 
 The value of $\Delta U^+$ is representative of the downward shift of the velocity profile in the wall-normal direction due to the increase of resistance induced by the wall roughness. 
 The value of the additive constant $C_s = 5.2$ is used in the present case.
  \begin{figure*}
\begin{center}
\subfloat[]{
\includegraphics[width=0.5\linewidth]{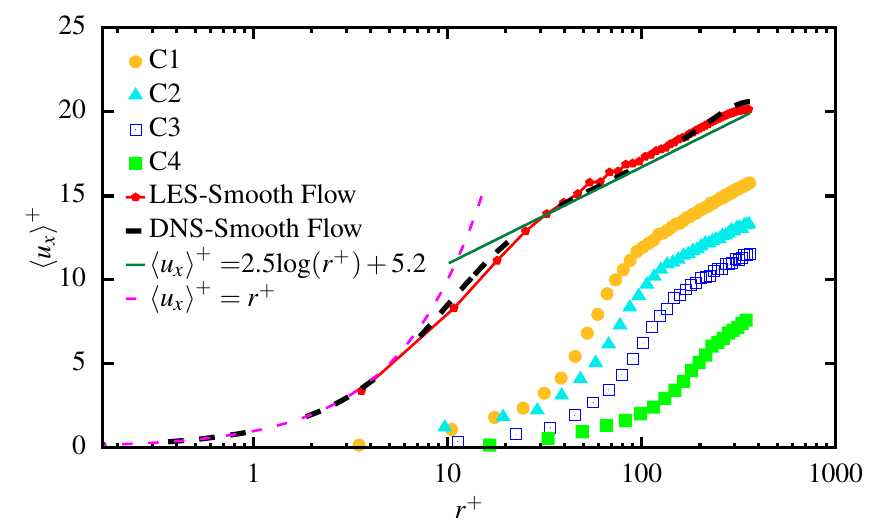}}
\subfloat[]{
\includegraphics[width=0.5\linewidth]{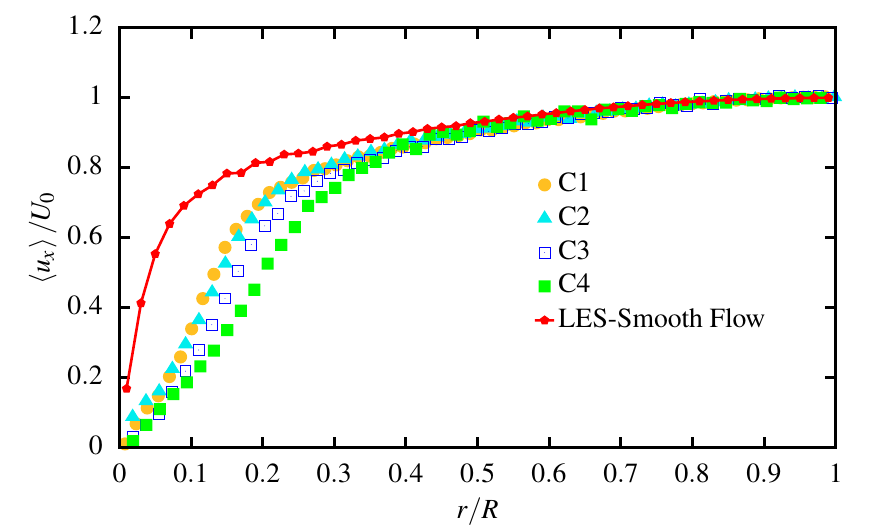}}

\subfloat[]{
\includegraphics[width=0.5\linewidth]{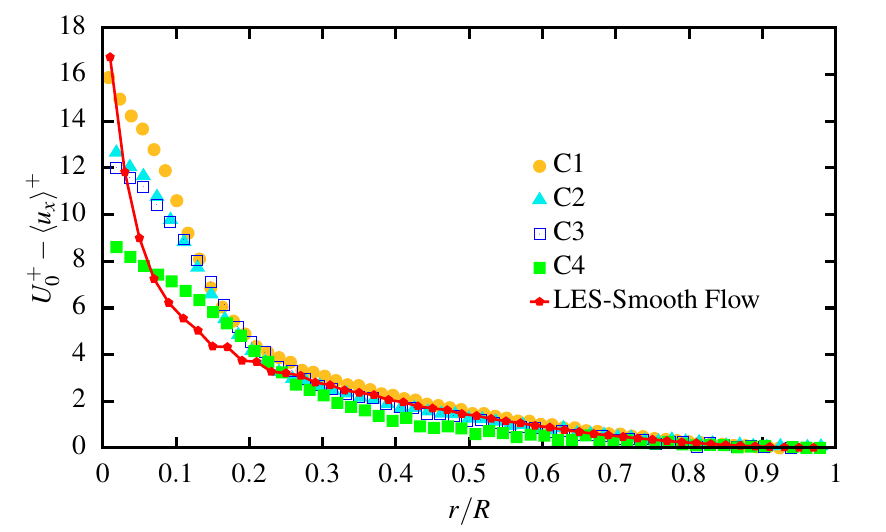}}
\caption{Comparison of LES results for rough and smooth pipe flows. (a) Mean streamwise velocity $\left <{u_x} \right > ^+$ as a function of $r^+$. The dashed black line presents DNS data at $Re_\tau = 361$. The dashed pink and green lines represent the universal behavior of the turbulent velocity profile. (b) Mean streamwise velocity normalized by centerline velocity as a function of $r/R$. (c) Mean streamwise defect velocity normalized by $u_\tau$ as a function of $r/R$.}
\label{fig: 1 Rough Surface mean flow profiles}
\end{center}
\end{figure*}

Firstly, in Fig. \ref{fig: flow visualizations}, we show a cross-sectional view of the mean streamwise velocity, normalized by $u_\tau$ for different rough surfaces, C1, C2, C3, and C4. Qualitatively speaking, when normalized by $u_\tau$, the downward shift of the streamwise velocity is clearly visible with increasing roughness. 
For detailed quantitative analysis, the non-dimensional numerical streamwise mean velocity profiles for different roughnesses are shown in semi-logarithmic plots in Fig. \ref{fig: 1 Rough Surface mean flow profiles} (\textit{a}). The non-dimensional wall distance, $r^+$, is measured from the plane at which the total drag acts, i.e., shifting the velocity profiles for rough surfaces by removing the negative velocities, a method frequently referred to as zero-plane displacement. In the same plot, we also provide the numerical results of mean streamwise velocity profiles for LES \citep{Himani2022} and DNS \citep{el2013direct} of smooth pipe flow at the same bulk Reynolds ($Re_b = 11,700$) for reference. We observe that the surface roughness shifts the logarithmic profile downwards; the shift is measured by $\Delta U^+$, indicating a shortage in momentum compared to smooth walls. The roughness heights of the considered surfaces are a significant fraction of the boundary layer thickness, resulting in large enough roughness functions. In literature, similar roughness functions are obtained for the sand-grain roughness at approximately the same Reynolds number, even though the roughness heights differ substantially \citep{flack2010review,kim2019piv,busse2017reynolds,thakkar2017surface,volino2007turbulence,connelly2006velocity,barros2019characteristics,busse2020influence}. This indicates that the roughness height alone is not sufficient to scale the momentum deficit resulting from surface roughness.

In Fig. \ref{fig: 1 Rough Surface mean flow profiles} (b), we also plot the mean velocity profiles in the outer units. We observe a significant departure of the rough pipe flow profiles from the one over the smooth surface close to the wall region and in the overlap region, $r/R < 0.4$. The more significant the roughness, the larger the shift from the mean flow profile of the smooth surface. In the outer region, irrespective of the roughness heights, all profiles collapse.  

For turbulent flow over smooth and rough surfaces, the wall similarity hypothesis, an extension of Townsend's Reynolds number similarity hypothesis, states that for sufficiently high $Re$ outside the viscous or roughness sublayer, the turbulent motion in the boundary layer is independent of the wall roughness and the viscosity, except for the wall shear stresses, $\tau_w$, and the boundary layer thickness, $\delta$ \cite{townsend1980structure,Raupach1991}. This means that the outer region is thought to be unaffected by roughness, as reported in review papers \cite{Raupach1991,jimenez2004turbulent,kadivar2021review}. 
 This further implies that the mean flow  for both smooth and rough walls obeys the universal velocity defect law in the overlap and outer regions of the boundary layer, which can be expressed as

\begin{equation}
U_0^+-\left< u_x\right>^+ = f\left(\dfrac{r}{R}\right).
\end{equation}
In an attempt to verify this statement's validity for AM rough surfaces, we plot the defect velocity of non-zero pressure gradient turbulent flow over different rough and smooth surfaces in Fig. \ref{fig: 1 Rough Surface mean flow profiles} (\textit{c}). The mean flow similarity can be observed by overlapping the smooth and rough walls when plotted in velocity-defect form for $r/R > 0.25$. This substantiates that the outer layer similarity may be maintained in the present 3D printed rough surfaces with relatively high roughness height, $\delta/R_a = 37-84$. Furthermore, the observation of the similarity for the rough surfaces whose roughness height is a significant fraction of the boundary layer thickness is in good accordance with the observations by experimental \citep{flack2007exp}, and DNS \citep{2005chan2015systematic} studies. Further evidence of wall similarity has also been provided in the review paper by \citet{flack2014rough}, evaluating the similarity hypothesis against the data of several studies with a range of roughness, from where similarity is expected to hold $(\delta/R_a = 110)$ to significantly rougher surfaces ($\delta/R_a=16$). Note that the range of roughness considered in the present study is relatively much higher than the ones considered by \citet{flack2007exp} but is comparable to the ones considered by Flack and Schultz \citet{flack2014rough}. 
\label{RS}
\subsection{Roughness function}
In the literature, the velocity shift and roughness function, as a result of wall similarity, are considered to be proportional to $\log k_s^+$ and are independent of wall distance \cite{cebeci2012analysis,saleh2005fully}. Several correlations have been suggested for roughness function based on the collapse, either to traditional Nikuradse data \cite{nikuradse1950laws} or Colebrook et al. \cite{colebrook1939correspondence} data. (For a fully rough regime, the Nikuradse data and Colebrook data should collapse, their difference is just in the transitional regime. For the present analysis, we have adopted the general form of the Colebrook-type roughness function\gs{,} given by
\begin{equation}
\Delta U^+ = 2.44\log{\left(1+0.26k_s^+\right)}.
\label{eq: collebrook}
\end{equation}
One can also find a correlation for the Nikuradse roughness function in \cite{andersson2020review}.
The value of $k_s^+$ is unknown a priori in our present study. The value of $k_s^+$ been computed by extracting the value of $\Delta U^+$ from Fig. \ref{fig: 1 Rough Surface mean flow profiles}(a) and utilizing the Eq. \ref{eq: collebrook}. An attempt to estimate the value of $k_s$ using the more recent correlation proposed by Marchis \textit{et al.} \cite{de2016large} based on $k_a$ and $ES$. But the results were inconsistent due to relatively small values of $ES$ for the present roughnesses and hence not included in this study. This point deserves a separate discussion and will be discussed in future investigations.

\begin{figure*}
\begin{center}
\subfloat[]{
\includegraphics[width=0.5\linewidth]{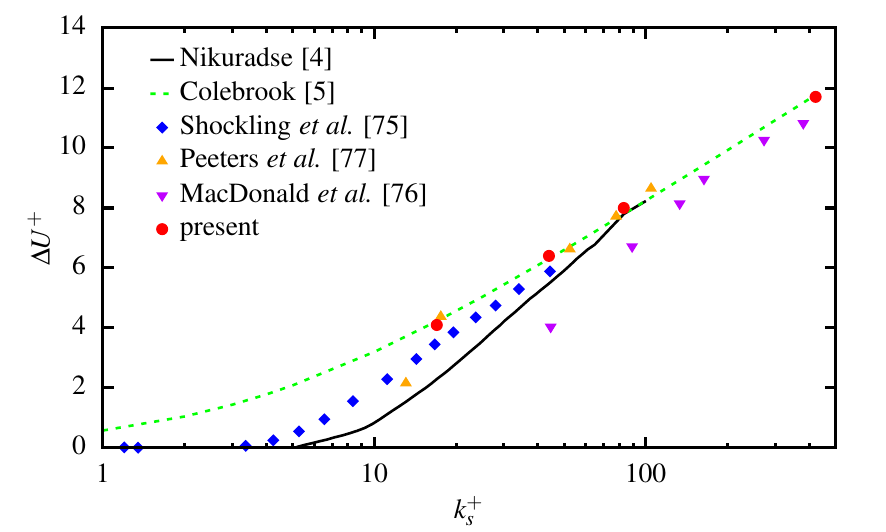}}
\subfloat[]{
\includegraphics[width=0.5\linewidth]{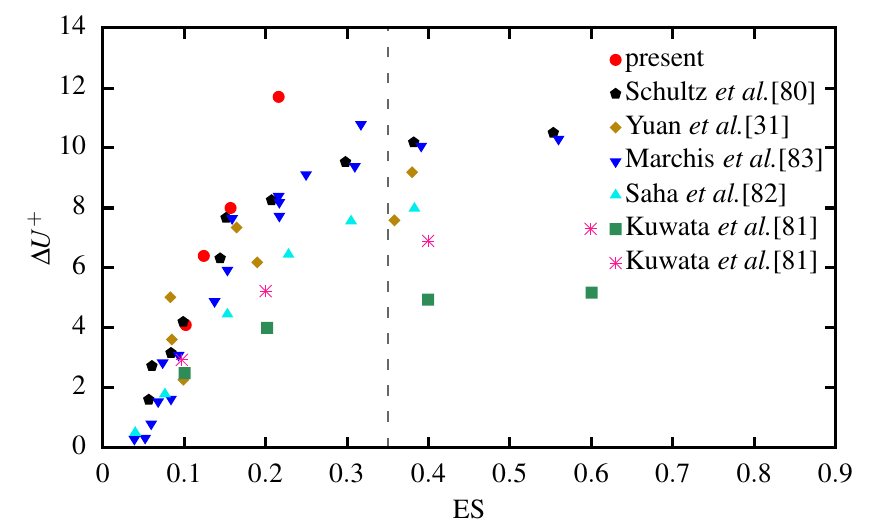}}
\caption{Roughness function as a function of (a) $k_s^+$ and (b) $ES$.}
\label{fig: 3 Rough Surface roughness function profiles}
\end{center}
\end{figure*}

In Fig. \ref{fig: 3 Rough Surface roughness function profiles}(a), we plot the relationship between the roughness function and the roughness Reynolds number, $k_s^+$, based on equivalent sand-grain roughness height, for a wide range of rough surfaces. For comparison, uniform sand-grain roughness results by Nikuradse \cite{nikuradse1950laws} and Colebrook \cite{colebrook1939correspondence}, rough-wall pipe flow results by Shockling et al. \cite{shockling2006roughness}, and rough-wall channel flow results by MacDonald et al. \cite{macdonald2019roughness} and Peeters and Sandham \cite{peeters2019turbulent} are presented.

Two flow regimes can be observed in Fig. \ref{fig: 3 Rough Surface roughness function profiles} (\textit{a}).
Firstly, for $k_s^+=17$, the flow becomes transitionally rough. In this regime, the viscosity can no longer dampen out the turbulent eddies created by the rough elements, and hence the form drag on the rough elements as well as the viscous drag. This indicates that the rough surface C1 with $k_s^+ \approx 17$ is roughly in the transition regime. However, this contrasts with some other works, where the surfaces display Nikuradse-type behavior, tending toward the hydraulically smooth flow regime at finite $k_s^+$ with slight inflectional behavior in the transition regime. This could be linked to the fact that in the present case, $k_s$ is based on Colebrook roughness function formulation rather than Nikuradse's, so a departure in the transition regime is seen.
Secondly, with a further increase of $k_s^+ > 17$, the roughness function reaches a linear asymptote, also known as a fully rough regime. For $k_s^+ > 17$, the C2, C3, and C4 roughnesses strictly follow the Colebrook-type roughness function. In this regime, the skin friction coefficient is independent of the Reynolds number, and form drag on the roughness elements is the dominant mechanism responsible for the momentum deficit. 

Surprisingly, the results indicate that the range of the transitionally rough regime is quite narrower, $k_s^+ < 25$. This contrasts with Nikuradse's that the fully rough regime exists for $k_s^+ \geq 70$ for mono-dispersed close-packed sand-grain roughness. However, \citet{ligrani1986structure} pointed out that $k_s^+$ signifying both the onset of roughness effects and the beginning of the fully rough regime depends strongly on the roughness type and uniformity. Their rough surfaces, consisting of uniform close-packed spheres, displayed a fairly narrow transitionally rough regime, $15<k_s^+ < 55$, while Nikuradse's uniform, close-packed sand grains, had a wider transitionally rough regime, $5< k_s^+ < 70$. 
Overall, the agreement between the present data sets and the literature is excellent. This concurs with the findings by \citet{schultz2003comparison}, who showed that the roughness function for a given roughness obtained through different means yields similar results. 

Usually, for dense roughness regimes, the roughness function does not scale with $k_s^+$. Hence, an additional factor, $ES$, is considered for scaling turbulent flows for all roughness regimes. In Fig. \ref{fig: 3 Rough Surface roughness function profiles} (\textit{b}),  we show the variation of the present roughness function, $\Delta U^+$, with $ES$. For comparison, we also plot different types of roughnesses available in the literature, i.e., systematically varied rough surfaces with positive and negative skewness \citep{kuwata2020direct}, smoothly corrugated walls with zero skewness \citep{saha2015scaling}, 2D and 3D roughness, as well as 2D waves \citep{de2016large}, irregular roughness with a range of positive and negative skewness \citep{yuan2014estimation}, and irregular surface roughness \citep{napoli2008effect}.

In Fig. \ref{fig: 3 Rough Surface roughness function profiles} (\textit{b}), for the dense roughness regime, $ES > 0.35$, $\Delta U^+$ decreases with increasing $ES$, which is attributed to a reduction in the Reynolds shear stress in a near wall region that is being pushed away from the rough wall. The comparison also shows that for irregular roughness, $\Delta U^+$ is more significant for the surfaces with positive $s_k$ than for the ones with negative $s_k$.
However, the calculated values of $ES$ are rather small ($< 0.35$) in comparison to the one observed in the literature, indicating that the present roughnesses might fall into a wavy roughness regime. Nevertheless, the results show the opposite: $\Delta U^+$ does not scale with $ES$. For $ES < 0.35$, the rough surfaces yield a higher value of $\Delta U^+$. This raises the question of whether these AM surfaces are in a wavy regime. To answer, we estimated the wavelength of the surface elevation, $\lambda = m_0/m_1$, where $m_0 = \int_0^{+\infty} E_{z'z'} dk$ and $m_1 = \int_0^{+\infty} k E_{z'z'} dk$ are zeroth- and first-order moments of power spectral density, $E_{z'z'}$, of the surface elevation and $k$ is the wave number. The obtained value of $\lambda$ is much smaller than $R$, which is a signature of sparsely packed and isolated roughness elements. This confirms that our AM roughnesses are not wavy but in a very sparsely packed roughness regime. We think that yielding a higher roughness function value, despite a smaller value of $ES$, could be a critical feature of AM rough surfaces. More investigations on AM surfaces are needed to validate this hypothesis.
\subsection{Turbulence intensities in the outer scaling}
 \begin{figure*}
\begin{center}
\subfloat[]{
\includegraphics[width=0.5\linewidth]
{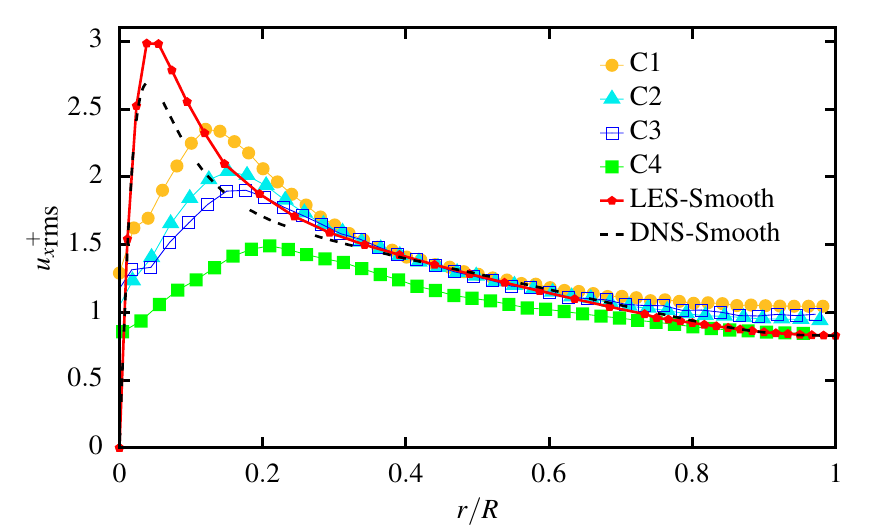}}
\subfloat[]{
\includegraphics[width=0.5\linewidth]
{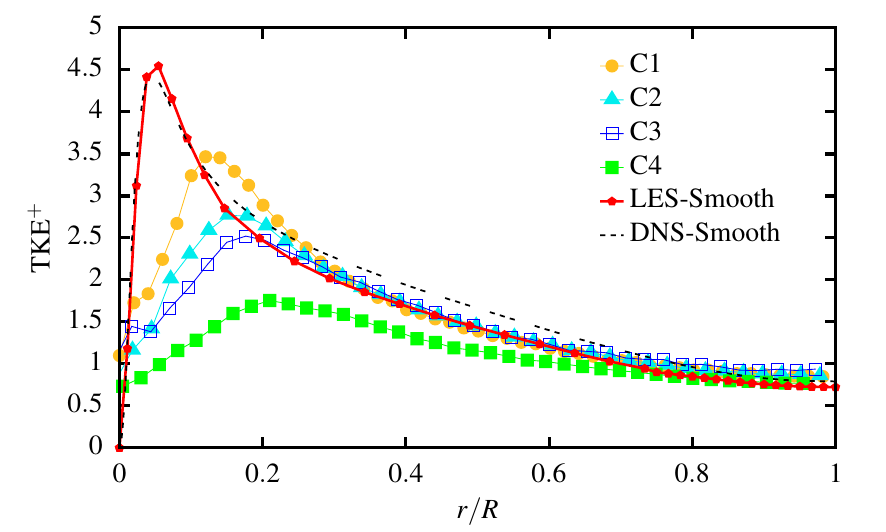}}

\subfloat[]{
\includegraphics[width=0.5\linewidth]
{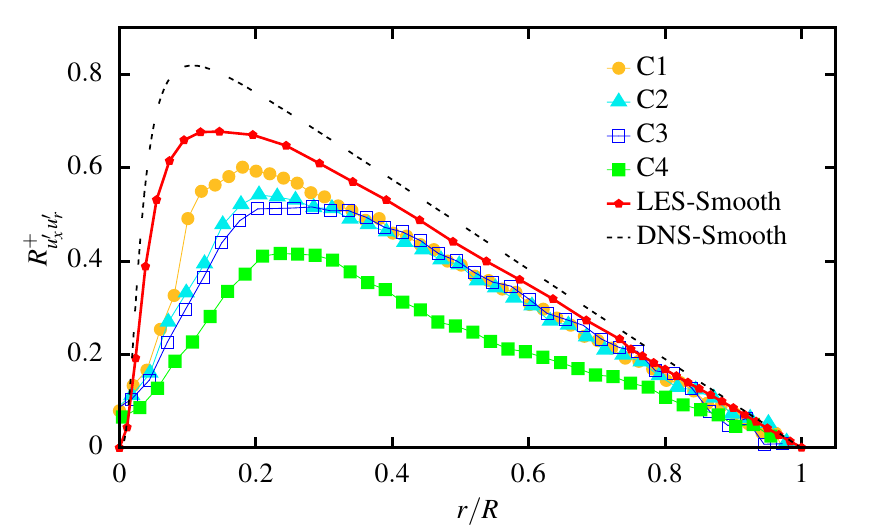}}
\caption{(a) Axial turbulence intensities, ${u_x}^+_{\text{rms}}$, (b) turbulent kinetic energy, TKE$^+$, and (c)  turbulent shear stresses, $R_{u_x' u_r'}^+$, for different rough surfaces, shown as a function of outer scaling, $r/R$. The results for smooth pipe flow for WALE-LES \cite{Himani2022} (red points) and DNS \cite{el2013direct} (dashed line), at $Re_b = 11,700$, are also plotted for reference.}
\label{fig: 4 Rough Surface turbulent intensity}
\end{center}
\end{figure*}
We have measured the intensity of the velocity fluctuations, a lowest-order indicator of the rough wall's influence on other flow parts. We show here the most important ones, streamwise velocity fluctuation profiles for different roughnesses in Fig. \ref{fig: 4 Rough Surface turbulent intensity} (\textit{a}), turbulent kinetic energy  in Fig. \ref{fig: 4 Rough Surface turbulent intensity} (\textit{b}), in outer scaling. The profiles for the wall-normal and azimuthal velocity fluctuations are not shown here, as the effect of roughness is weaker on these quantities. For comparison, we also show the profiles for smooth pipe flows at the same bulk Reynolds number, $Re_b = 11,700$, obtained from LES \citep{Himani2022} and DNS \citep{el2013direct}. 

As $k_s^+$ increases and the form drag dominates over the viscous scale, the near-wall peak in the streamwise velocity fluctuations, ${u_x}_{rms}$, disappears, and the profile develops a maximum in the logarithmic layer around $r/R = 0.05-0.2 $. This outer peak is probably related to the plateau found in that region for high Reynolds number flows over smooth walls \citep{de2000reynolds}.  Also, in the outer region, all the present roughnesses except C4 collapse very well for $r/R > 0.3$, with reasonable numerical uncertainty. These small mismatches between the smooth pipe flow and rough surface flow in the outer scale could be attributed to the statistical averaging. Furthermore, for the C4 case, the effect of flow blockage may be significant, resulting in higher form drag, and an eventually slight departure from Townsend's similarity is observed.
However, overall these results support Townsend's Reynolds number similarity hypothesis that the flow is insensitive to the wall boundary condition except for the role it plays in setting the outer scale and velocity scales.

In addition, we also plot the turbulent Reynolds shear stress profiles in Fig. \ref{fig: 4 Rough Surface turbulent intensity} (\textit{c}). The profiles are normalized by $u_\tau^2$. As demonstrated earlier, with an increase in $k_s^+$ from C1 to C4 cases, the peak of shear stresses not only diminishes but also shifts outward, thanks to the high form drag contribution. A pretty good collapse of all rough surfaces except C4 for $r/R >  0.3$ has been observed. The results are relatively close to the LES smooth-wall case with little uncertainties, which could be linked to the approximation of friction velocity estimate. This once again demonstrates the outer scale similarity, as noticed in previous works. Note that, in the C4 case, where $R/R_a < 50$, while for C1 and C3 cases, $R/R_a > 50$ which belong to the roughness regime, as mentioned by \citet{jimenez2007recent}. This shows that there is probably a transition between C3 and C4 rough surfaces where AM roughnesses differ quite a lot in terms of their behavior and could drastically influence heat transfer properties.

\begin{figure*}
\begin{center}
\subfloat[]{
\includegraphics[width=0.5\linewidth]{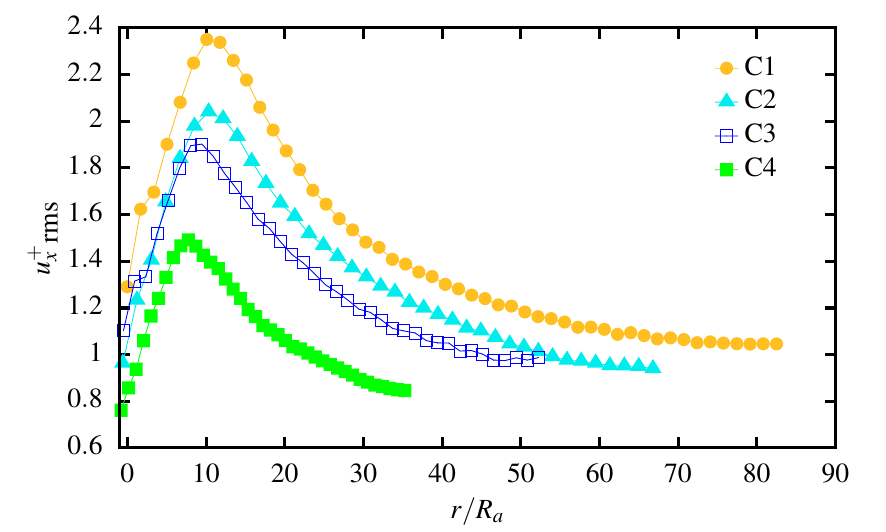}}
\subfloat[]{
\includegraphics[width=0.5\linewidth]{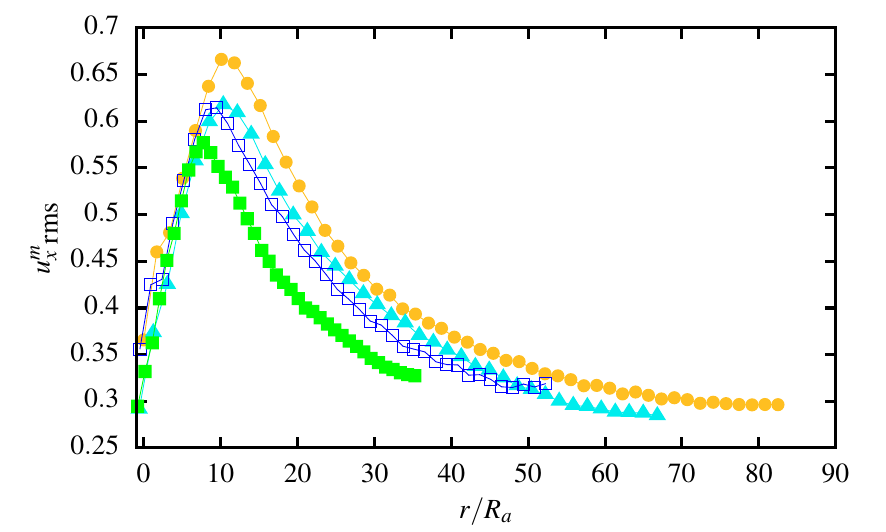}}

\caption{Streamwise turbulence velocity fluctuations, ${u_x}_{\text{rms}}$, shown as a function of inner scaling based on average roughness height, $r/R_a$, for different rough surfaces, and normalized by (a) $u_\tau$ and (b) $\sqrt{u_\tau U_{\text{0}}}$, a new mixed scaling introduced here.}
\label{fig: 4 Rough Surface turbulent intensity in inner scale units}
\end{center}
\end{figure*}

We are further interested in scaling streamwise velocity fluctuation profiles with inner scale, i.e., $r/R_a$, based on the average roughness height, $R_a$. We show the comparison of ${u_x}_{rms}$, normalized by $u_\tau$, as a function of $r/R_a$ in Fig. \ref{fig: 4 Rough Surface turbulent intensity in inner scale units} (\textit{a}).  
With increased $k_s^+$ values, the peak value close to the wall diminishes and shifts inwards.  All profiles develop a maximum in the range of $r/R_a = 8-10$. 
This inward shift is attributed to the yielding of the higher value of $u_\tau$, despite the relatively small value of the bulk Reynolds number, a typical characteristic of roughness. Nevertheless, one point that stands out here is that the roughness profiles do not fully collapse in the inner region, which implies that $r/R_a$ is not an accurate scaling for roughness. 
 This requires introducing another scaling related to roughness in the normalization of ${u_x}_{rms}$. Following the proposition by \citet{de2000reynolds}, we introduced a new scaling, referred to as mixed scaling from hereon, $\left < ...\right >^m = \left < ...\right >/(\sqrt{u_\tau U_0})$, in the normalization of ${u_x}_{rms}$, for which results are shown in Fig. \ref{fig: 4 Rough Surface turbulent intensity in inner scale units} (\textit{b}). For AM roughnesses, this scaling also works reasonably well, but there is an additional decrease in the peak value with increasing $k_s^+$, probably linked to the slight over-approximation of $u_\tau$. The presence of the freestream velocity, $U_0$, in the scaling of the streamwise fluctuations is usually associated with the effect of the inactive motions postulated by \citet{townsend1980structure}. However, the details of how they affect the streamwise fluctuations away from the wall layer are poorly understood.
 
\begin{figure}
\begin{center}
\includegraphics[width=1\linewidth]{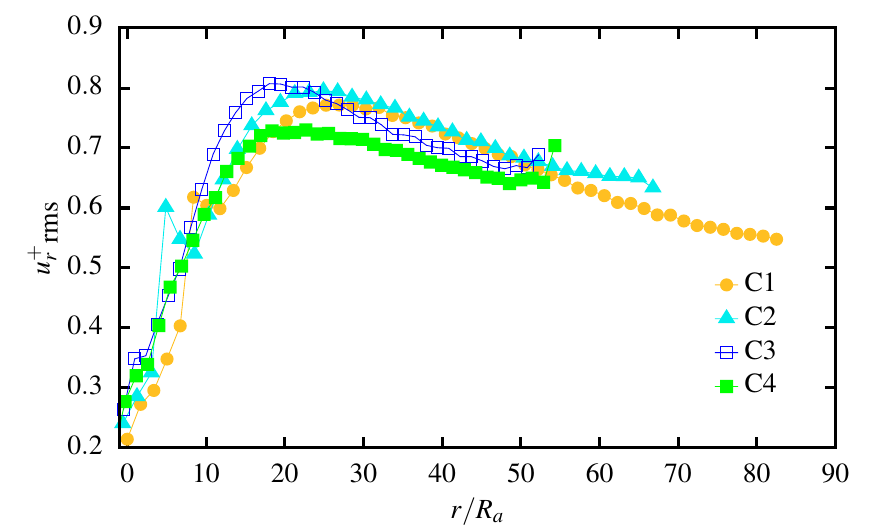}
\caption{Profiles of root-mean-square of radial velocity fluctuations, ${u_r^+}_{\text{rms}}$, normalized by $u_\tau$ are shown as functions of inner scaling based on average roughness height, $r/R_a$, for different rough surfaces.}
\label{fig: 4 Rough Surface wall-normal intensity}
\end{center}
\end{figure}

Figure \ref{fig: 4 Rough Surface wall-normal intensity} shows the profile of radial velocity fluctuations as a function of the inner scale, based on $R_a$. The roughness effect is weaker for this quantity, as the profiles are pretty close to each other, irrespective of the roughness. Consistently with previous results, the scaling with $R_a$ shows a good collapse in the inner-wall region for all roughnesses. The slight discrepancy could be attributed to the statistical averaging uncertainty. These results further show that there is no need to introduce mixed scaling, which implies the negligible contribution of the inactive motions to this quantity, as postulated by \citet{townsend1980structure}.

\begin{figure}
\begin{center}
\includegraphics[width=1\linewidth]{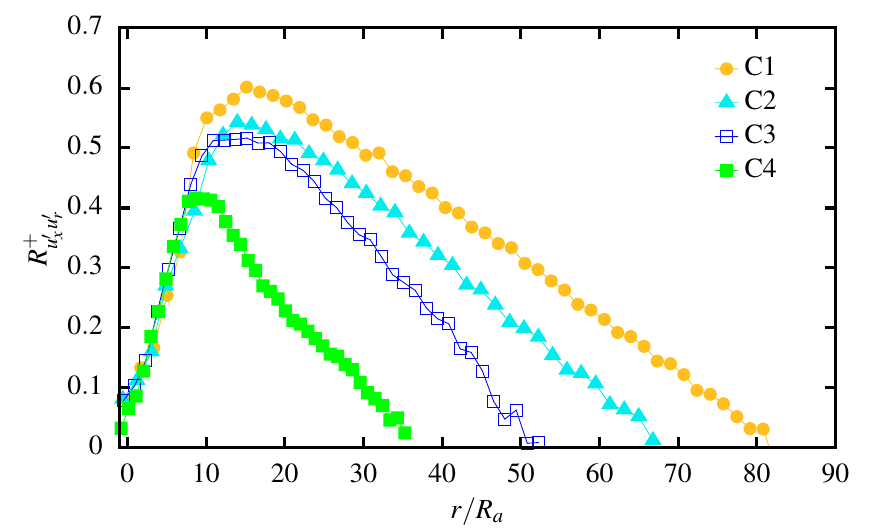}
\caption{Turbulence shear stresses, $R^+_{u_x'u_r'}$, normalized by $u_\tau^2$ are shown as functions of inner scaling based on average roughness height, $r/R_a$, for different rough surfaces.}
\label{fig: 4 Rough Surface shear stresses}
\end{center}
\end{figure}

\begin{figure*}
\begin{center}
\subfloat[]{
\includegraphics[width=0.5\linewidth]{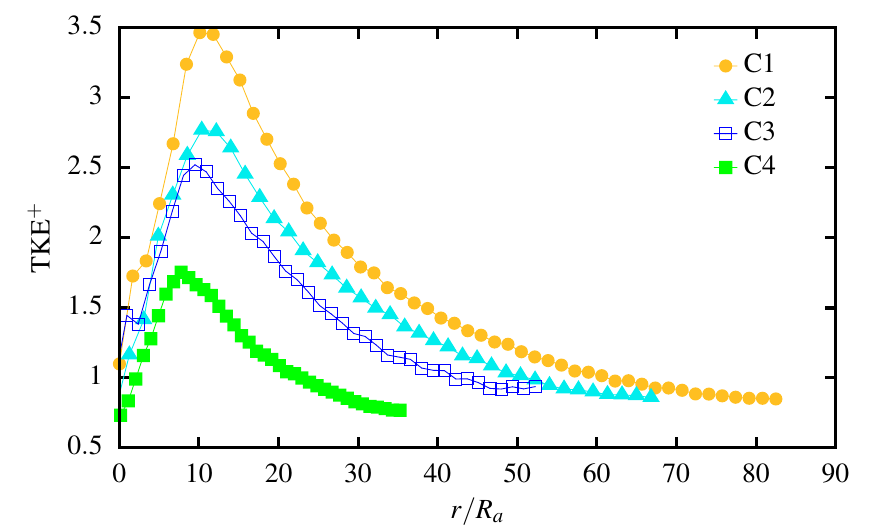}}
\subfloat[]{
\includegraphics[width=0.5\linewidth]{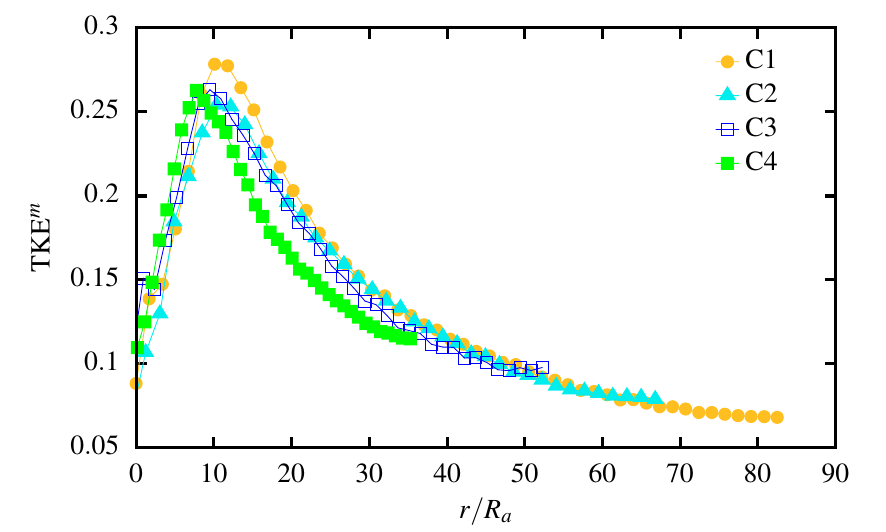}}
\caption{
Turbulence kinetic energy, TKE, expressed in terms of inner scaling based on average roughness height, $r/R_a$, for different rough surfaces, normalized by (a) $u_\tau^2$ and (b) $u_\tau U_0$, a new mixed scaling introduced here.}
\label{fig: 4 Rough Surface turbulent kinetic energy}
\end{center}
\end{figure*}

Similarly, the turbulent Reynolds shear stresses normalized by the friction velocity squared as a function of $r/R_a$ are presented in Fig. \ref{fig: 4 Rough Surface shear stresses}. As expected, the shear stress scales well in the inner wall region when plotted as a function of $r/R_a$. The peak value for different roughnesses increases with decreasing $k_s^+$ due to the scaling involved in $R^+_{u_x'u_r'}$. Indeed, the larger the roughness, the larger the form drag and hence $u_\tau$. To end this discussion, we finally show the plots of turbulent kinetic energy for different scalings in Fig. \ref{fig: 4 Rough Surface turbulent kinetic energy}. The turbulent kinetic energy is mainly dominated by the streamwise velocity and shows very similar behavior. The results clearly scale very well in the inner wall region once the mixed scaling is introduced.

\begin{figure*}
\begin{center}
\subfloat[]{
\includegraphics[width=0.5\linewidth]{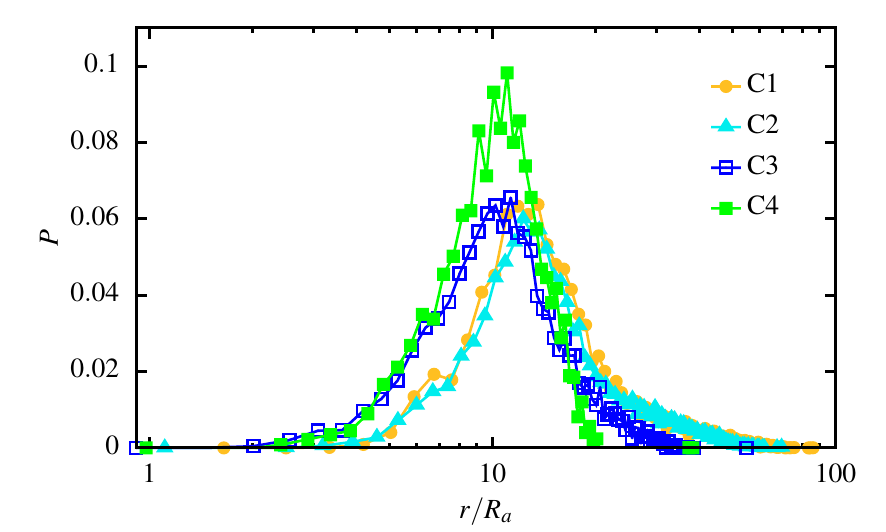}}
\subfloat[]{
\includegraphics[width=0.5\linewidth]{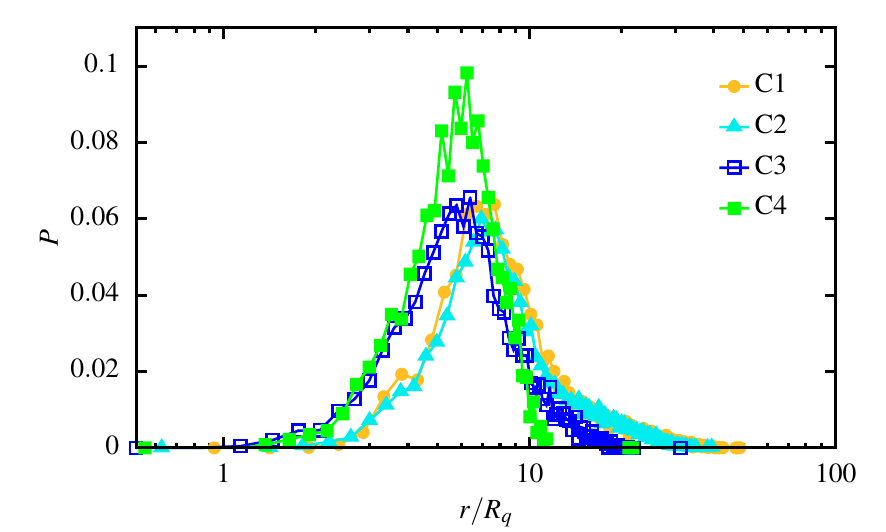}}
\caption{Turbulent kinetic energy production as a function of inner scale units based on (a) average roughness, $r/R_a$, (b) root-mean-square roughness scale units, $r/R_q$.
Here, $P=-R_{u_x'u_r'}\partial\left<u_x^+\right>/\partial r^+$ is evaluated using the law of the wall-wake for the mean velocity, for zero-pressure gradient boundary layers.}
\label{fig: 7 Turbulent kinetic energy production}
\end{center}
\end{figure*}
\subsection{Turbulent kinetic energy production}
In both DNS and experiments of smooth-wall flows, it has been shown that for low Reynolds numbers, the peak of kinetic energy production mainly occurs within the viscous buffer layer region, at a wall-normal distance, $r^+$, of roughly $12$. However, at high Reynolds numbers, the significant contribution to the bulk turbulence production comes from the logarithmic region. It raises a similar interesting question in the case of AM roughnesses. Therefore, in the present simulations, we looked into the contribution to the bulk turbulence production. 
In Fig. \ref{fig: 7 Turbulent kinetic energy production}, we show the production term, $P=-R_{u_x'u_r'}\partial\left<u_x^+\right>/\partial r^+$, on a semi-logarithmic scale. In Fig. \ref{fig: 7 Turbulent kinetic energy production}(a-b), we show the results for the production term but scaled with the average roughness height, $R_a$, and the root-mean-square value of roughness height, $R_q$. Irrespective of the scaling factor, we observe that all curves collapse into one with slight uncertainties. In comparison to smooth-wall pipe flow, the maximum value of P for the present roughnesses is smaller and roughly one half of smooth pipe flows \cite{Himani2022}. Also, the peak value increases in the inner layer and shifts inward for all roughnesses. The results depict that the primary contribution to bulk production comes from the near-wall region at low $k_s^+$, and the logarithmic region dominates at adequately higher $k_s^+$.  
\subsection{Rough surface joint probability density functions}
 \label{JPDF}
In this section, we analyze the axial and radial velocity fluctuations for a fixed value of $r^+$ using the quadrant analysis approach \citep{wallace1972wall}, by the desire to gain insight into the near-wall streaks. 
We divided the ($u_x', u_r'$) plane into four quadrants corresponding to different sign combinations of the velocity fluctuations: two quadrants contributing negatively to $\left< u_x'u_r'\right>$, attributed to ejections (second quadrant, $Q_2$; $u_x'<0$, $u_r'>0$) and sweeps (fourth quadrant, $Q_4$; $u_x'>0$, $u_r'<0$), and two quadrants contributing positively to $\left< u_x'u_r'\right>$, owing to outward (first quadrant, $Q_1$; $u_x'>0$, $u_r'>0$) and inward (third quadrant, $Q_3$; $u_x'<0$, $u_r'<0$) interactions. 

 \begin{figure*}
\begin{center}
\subfloat[Smooth surface]{
\includegraphics[width=0.32\linewidth]{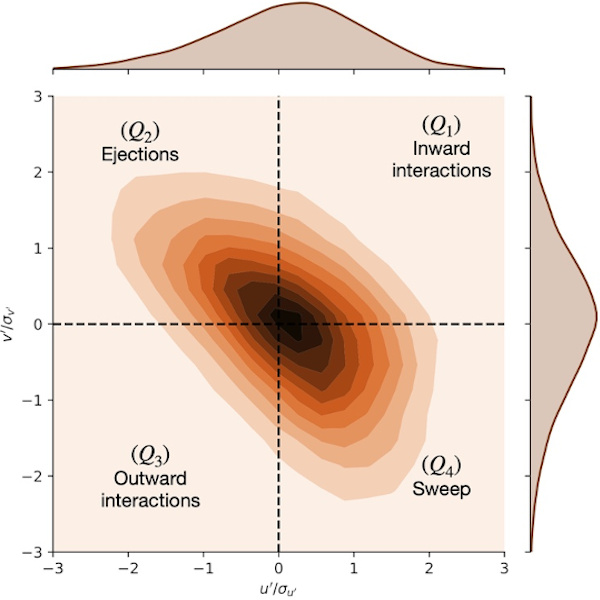}}
\subfloat[C1]{
\includegraphics[width=0.33\linewidth]{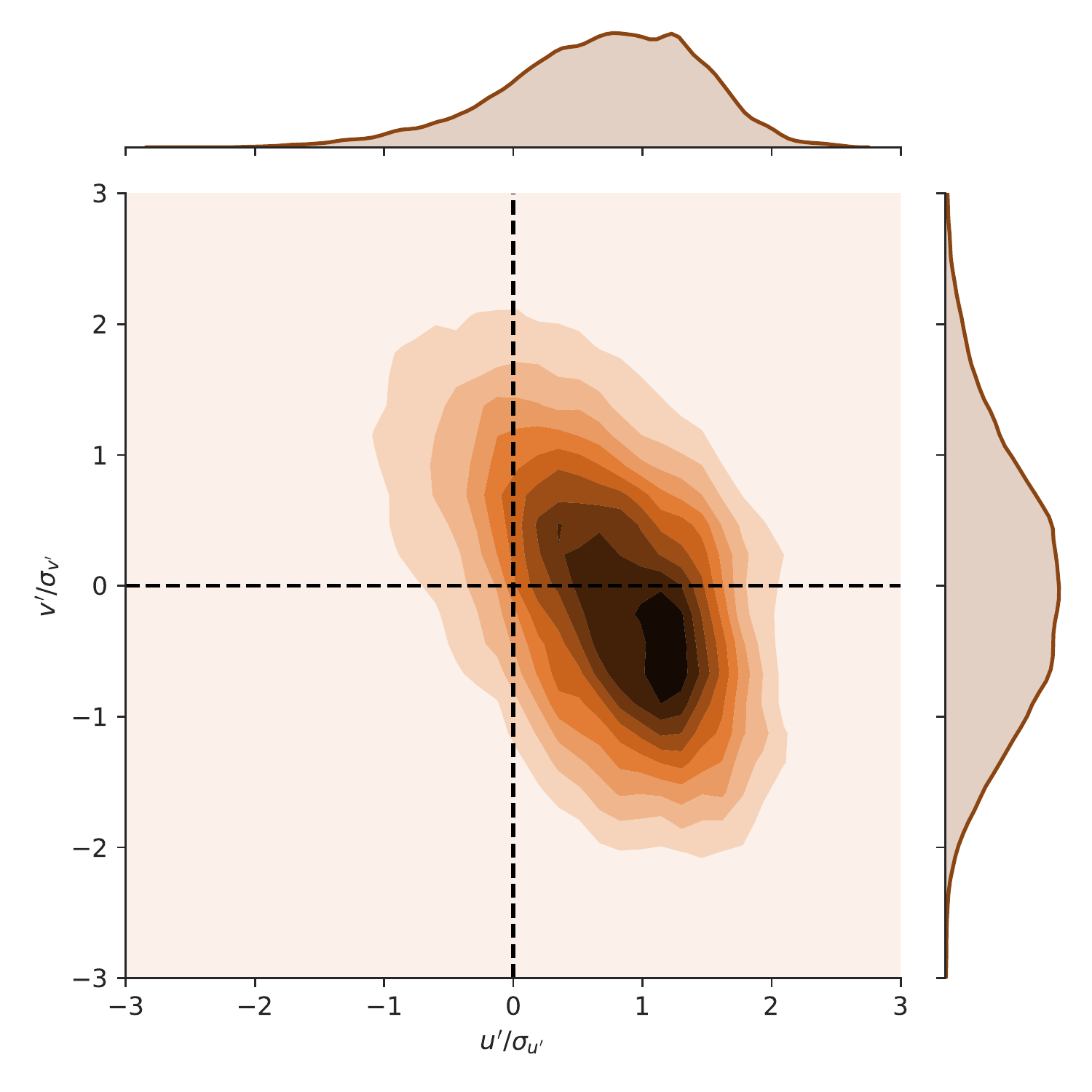}}
\subfloat[C2]{
\includegraphics[width=0.33\linewidth]{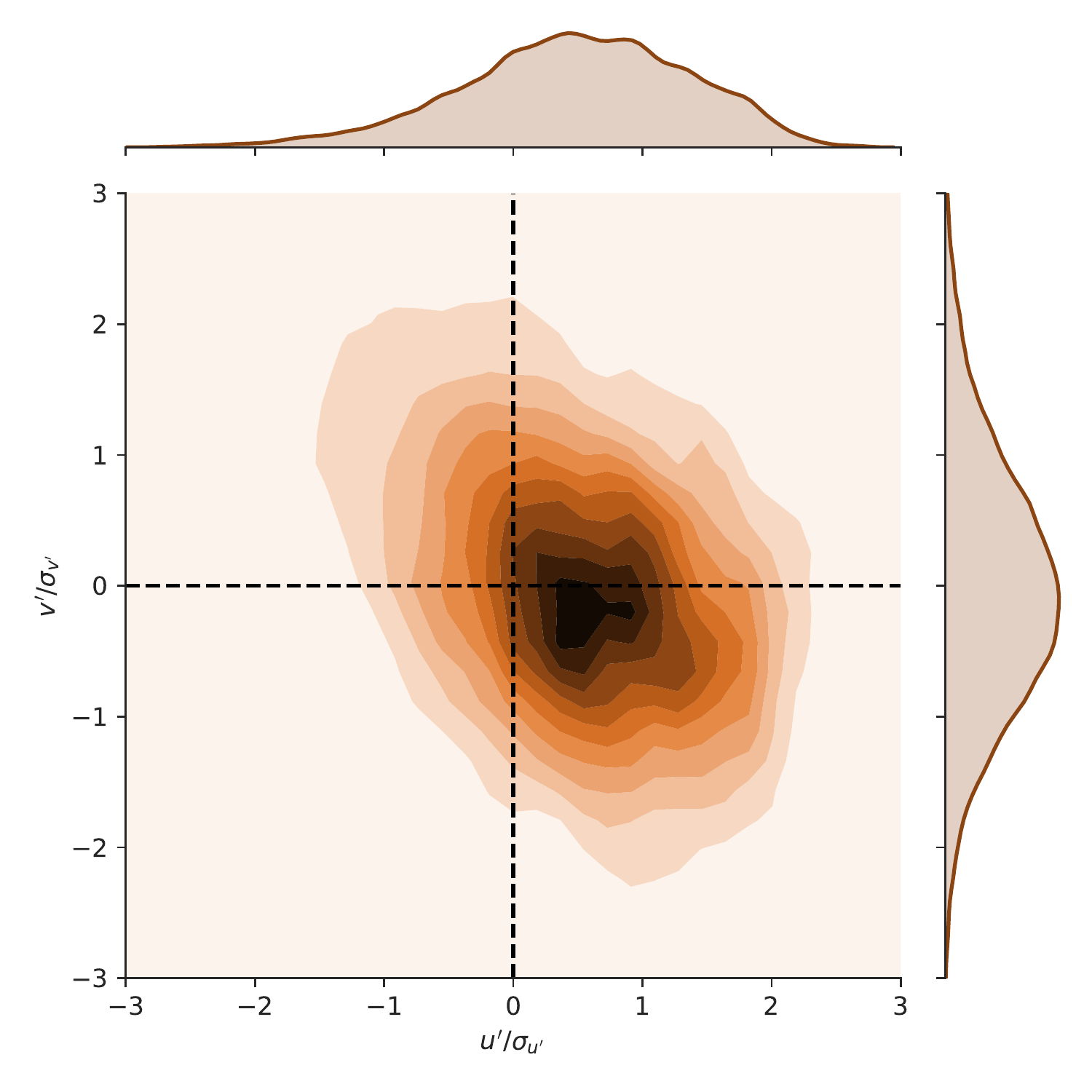}}

\subfloat[C3]{
\includegraphics[width=0.33\linewidth]{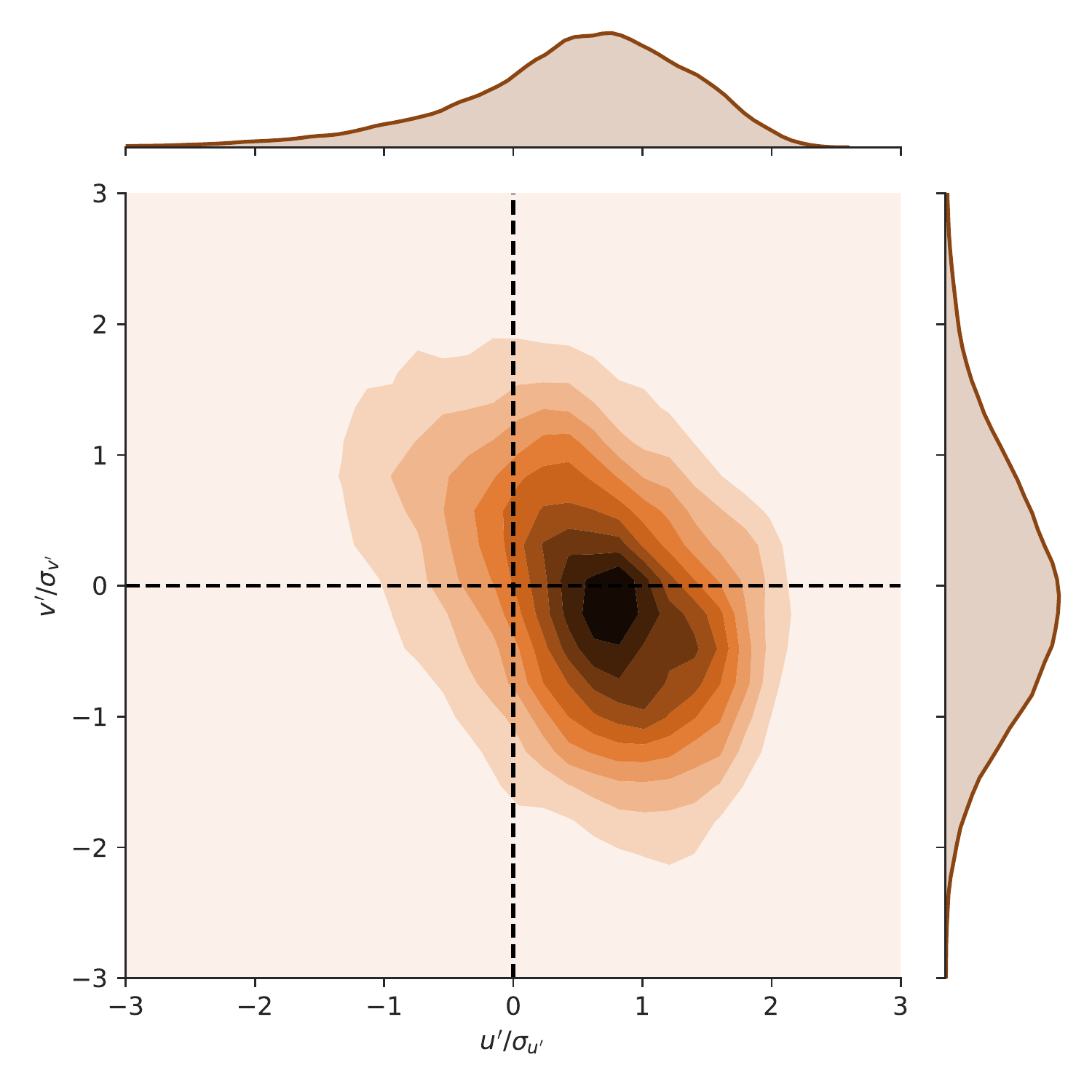}}
\subfloat[C4]{
\includegraphics[width=0.33\linewidth]{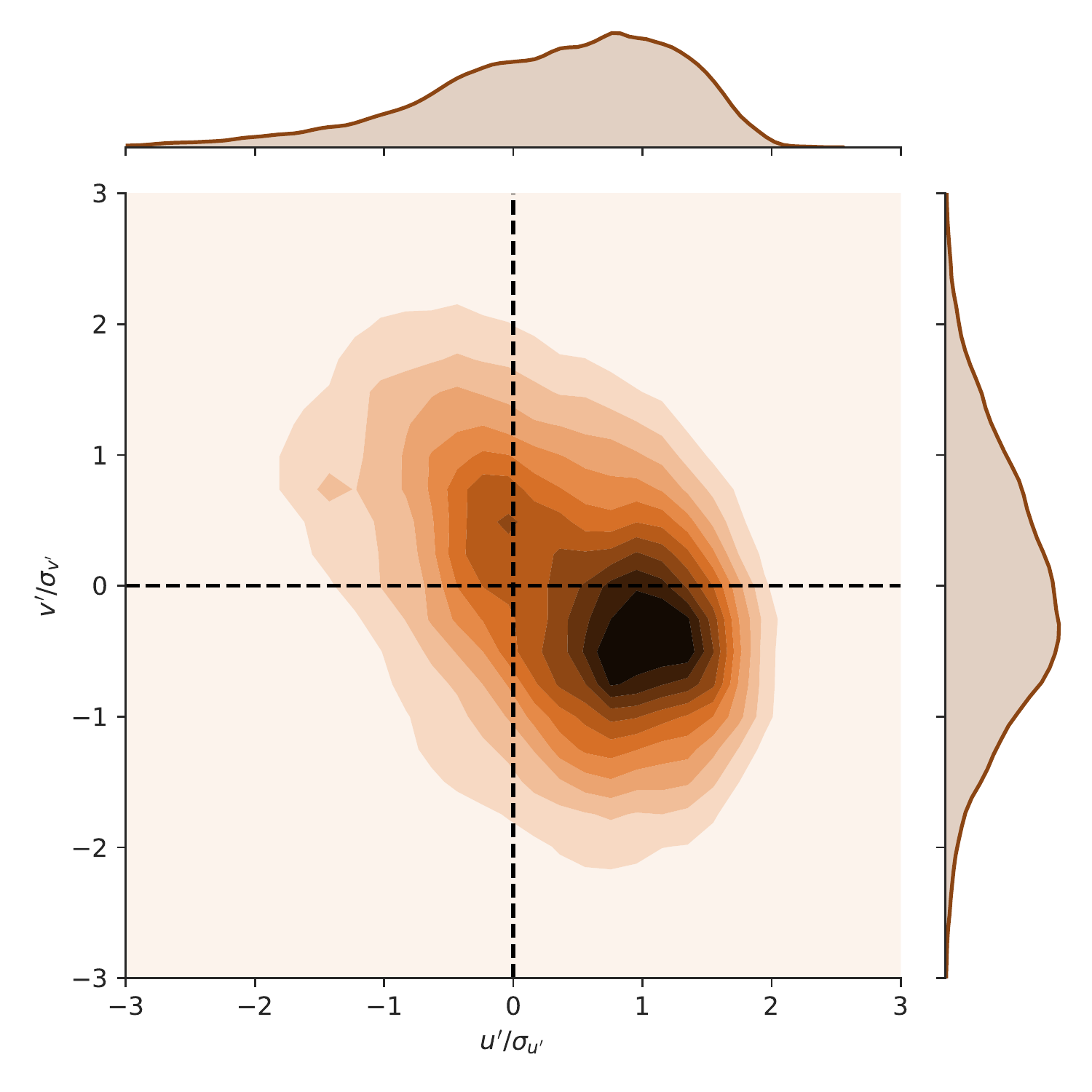}}
\caption{Typical quadrant plots (JPDF) are shown for illustration purposes between axial velocity fluctuations, $u_x'$, and radial velocity fluctuations, $u_r'$, normalized by their root-mean-squares ($\sigma_{i'}$) for different roughness Reynolds numbers at $r^+ = 3\sqrt{k_s^+}$: (a) smooth surface, (b) C1, (c) C2, (d) C3, and (e) C4. For each plot, the PDF of $u_x'/\sigma_{u_x'}$ and $u_r'/\sigma_{u_r'}$ are shown on the top and right corners, respectively.}
\label{fig: 3 JPDS for different fixed yplus using contours}
\end{center}
\end{figure*}
 
In Fig. \ref{fig: 3 JPDS for different fixed yplus using contours}, we present the contour plots of the Joint Probability Density Functions (JPDF) of velocity fluctuations, $J(u_x',u_r')$. The JPDF are taken at the closest location to $r^+ = 3\sqrt{k_s^+}$ (indicative of the middle of the logarithmic region) for different roughnesses, normalized by their local standard deviations, $\sigma_{u_x'}$, and $\sigma_{u_r'}$. For comparison, the JPDF of smooth-wall turbulent pipe flow is also shown. For all cases, the isocontours of the JPDF exhibit an elliptical shape whose main axis crosses quadrants $Q_2$ and $Q_4$. For the smooth surface case, the JPDF maximum value is located in quadrant $Q_4$, and a long tail is observed in quadrant $Q_2$. This suggests the highest number of $\left\{u_x'u_r'(t) < 0\right\}$ events, with, in particular, $Q_2$ (ejections) being characterized by more extreme events, i.e., with higher values of $|u_x'u_r'(t)|$. Overall the
smooth-wall results show a higher degree of homogeneity. In contrast, ejection and sweep contributions for rough-wall flow exhibit differences in the form of localized regions of more intense events, and $Q_4$ (sweeps) are characterized by more extreme events. 
In Fig. \ref{fig: 3 JPDS for different fixed yplus using contours}, for each plot, normalized Probability Density Functions (PDF) are also shown on the top ($u_x'$) and right ($u_r'$) corners, respectively. Notably, the PDF of $u_x'$ and $u_r'$ are slightly positively and negatively skewed for smooth surfaces, respectively. On the other hand, for all roughnesses, we observe that the PDFs of $u_x'$ and $u_r'$ are largely positively skewed and slightly positively skewed, respectively. This means high-speed streaks are dominant structures in rough surface flows compared to smooth surface flows. Qualitatively, the roughness has a negligible impact on the behavior of the results and delivers a similar message. The contribution of ejections and sweeping events has also been computed quantitatively, but it requires a separate long discussion and hence is not included in this paper.
\subsection{Roughness-induced secondary flow structures}
One intriguing and practically significant manifestation of turbulence in wall-bounded flows is the ability to induce the so-called secondary flows, that is, systematized fluid motions in the plane perpendicular to the streamwise direction. 
In turbulent regimes, the average velocity field contains non-zero transverse components, and secondary flow occurs in non-circular cross-section ducts related to inhomogeneities of the Reynolds stresses. 
The spanwise inhomogeneity produces distinct gradients in the turbulent Reynolds stress distribution and alters the local flow properties. 
The development of secondary flow motions can also occur on a plane or symmetrical wall-bounded flows if a local spanwise heterogeneity in wall conditions exists, e.g., due to surface roughness. 
 \begin{figure*}
\begin{center}
\subfloat[]{
\includegraphics[width=0.5\linewidth]{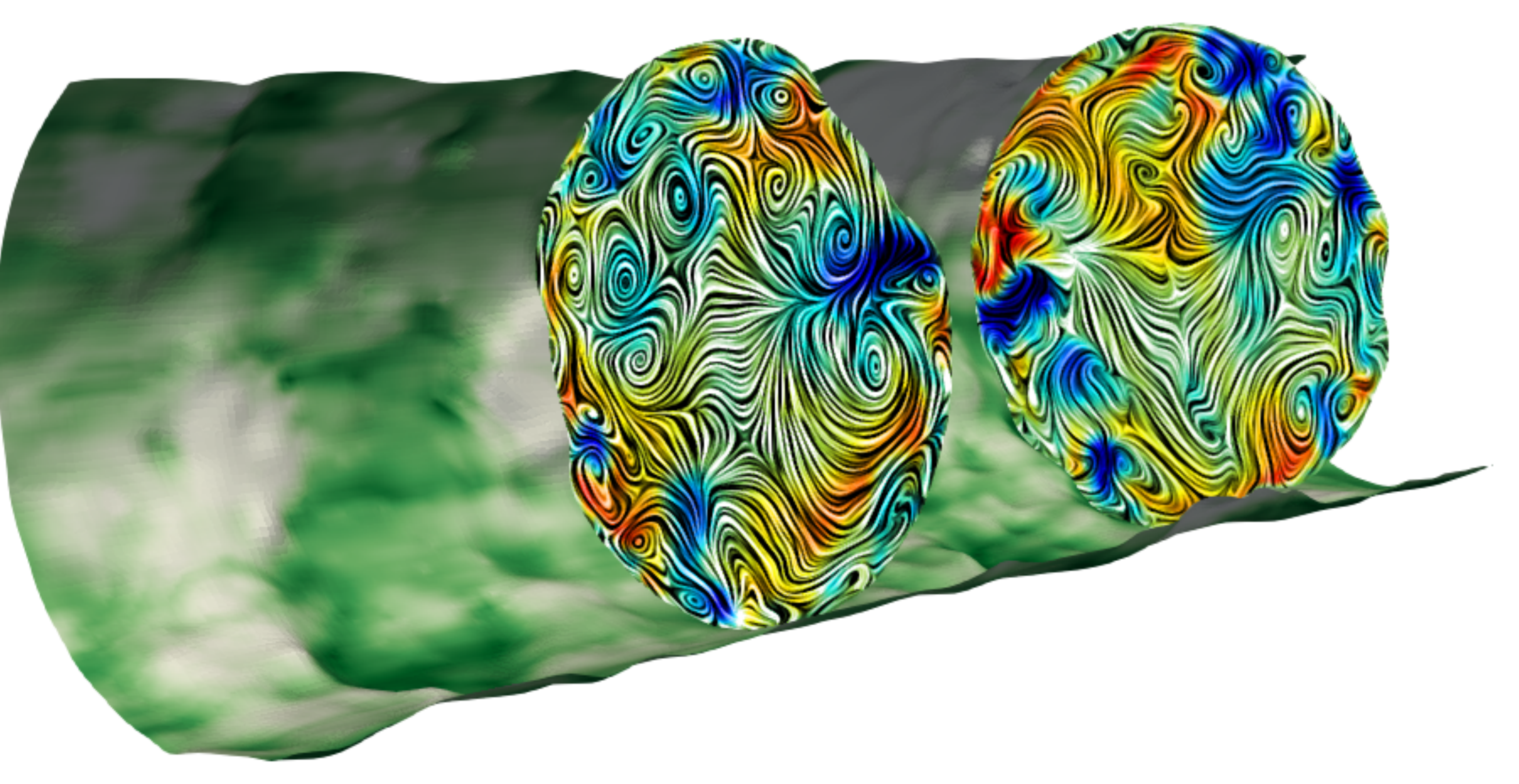}
         \label{fig:FlucLIC02d}}        
\subfloat[]{
\includegraphics[width=0.5\linewidth]{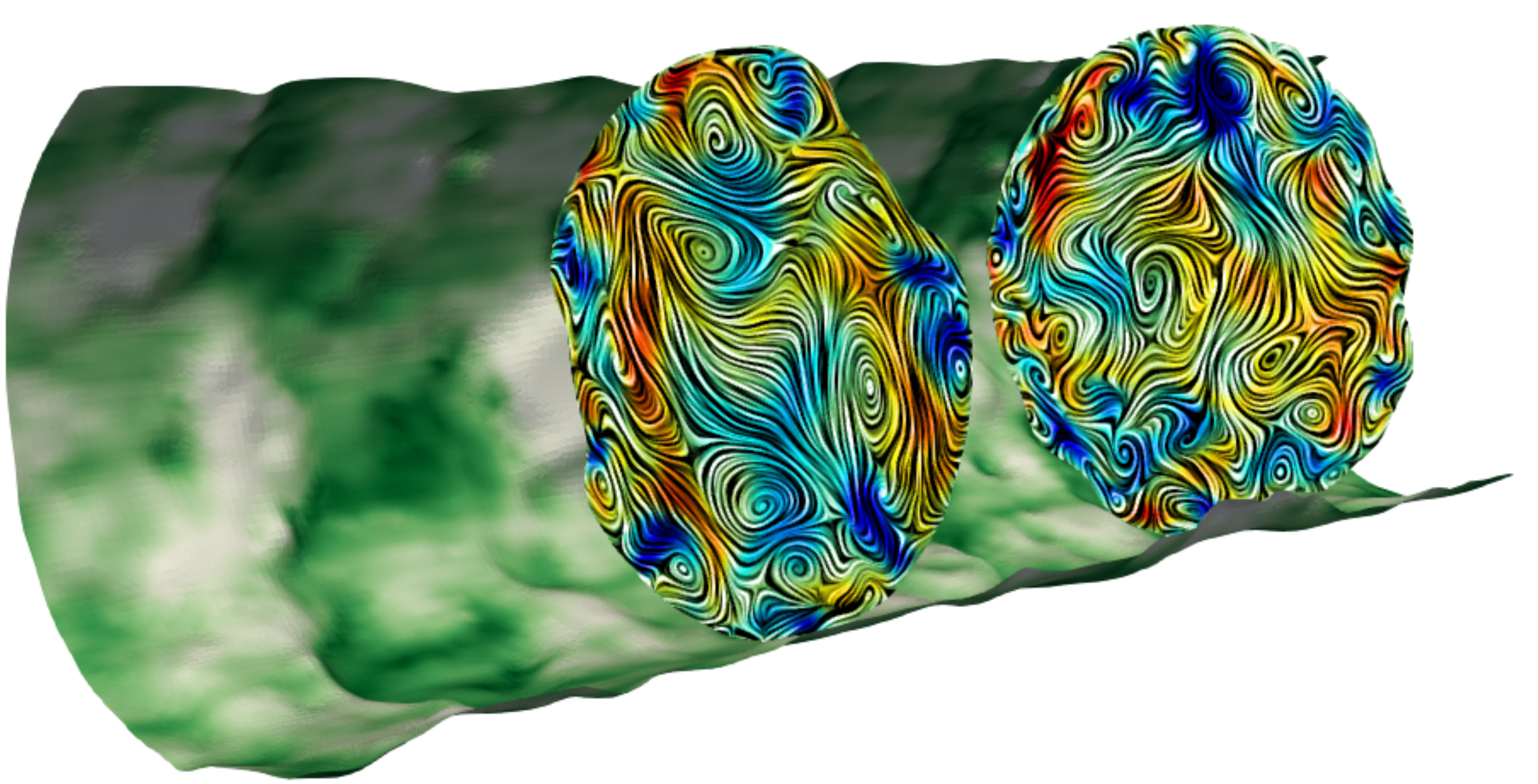}
         \label{fig:FlucLIC025d}}

\subfloat[]{
\includegraphics[width=0.5\linewidth]{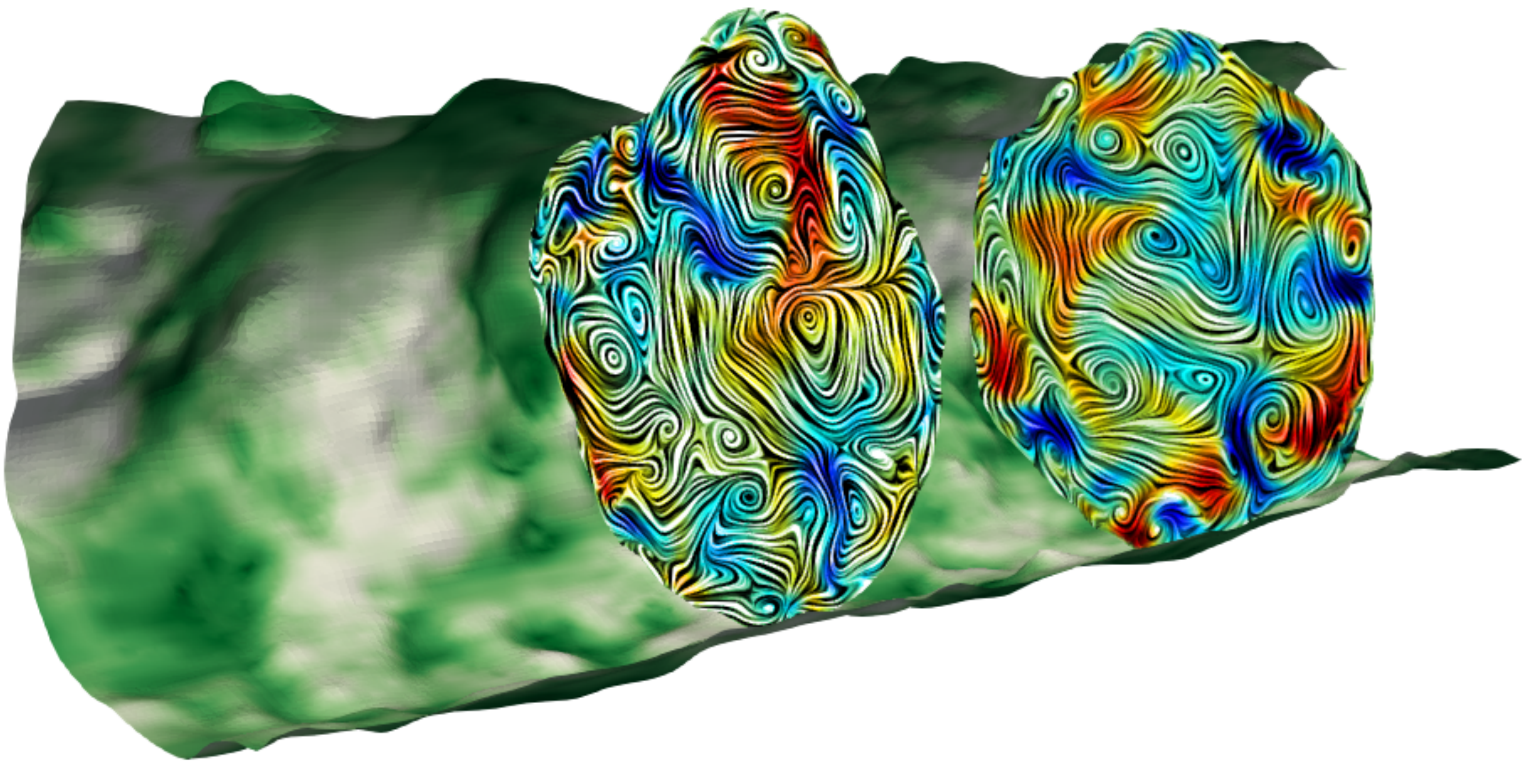}
         \label{fig:FlucLIC033d}}   
\subfloat[]{
\includegraphics[width=0.5\linewidth]{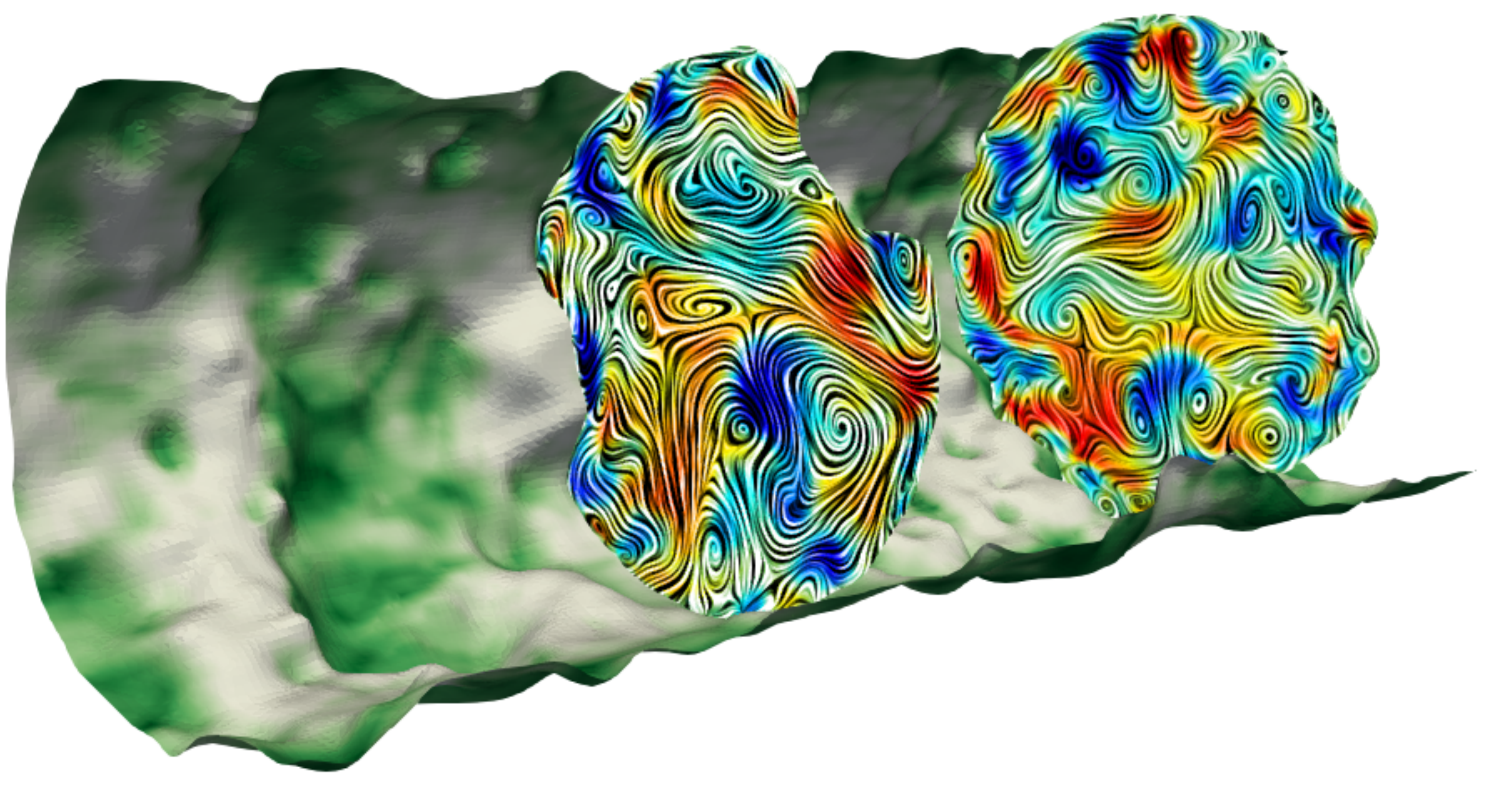}
         \label{fig:FlucLIC05d}}
\caption{Cross-sectional view of the instantaneous streamwise velocity fluctuations field for different rough surfaces: \subref{fig:FlucLIC02d} C1, \subref{fig:FlucLIC025d} C2, \subref{fig:FlucLIC033d} C3, and \subref{fig:FlucLIC05d} C4. Streamlines from the instantaneous flow field show the flow development and vortical flow patterns attributed to wall roughness. In all plots, the wall is colored with the friction velocity, $u_\tau$.}
\label{fig: FlucSTLIC}
\end{center}
\end{figure*}
 \begin{figure*}
\begin{center}
\subfloat[]{
\includegraphics[width=0.5\linewidth]{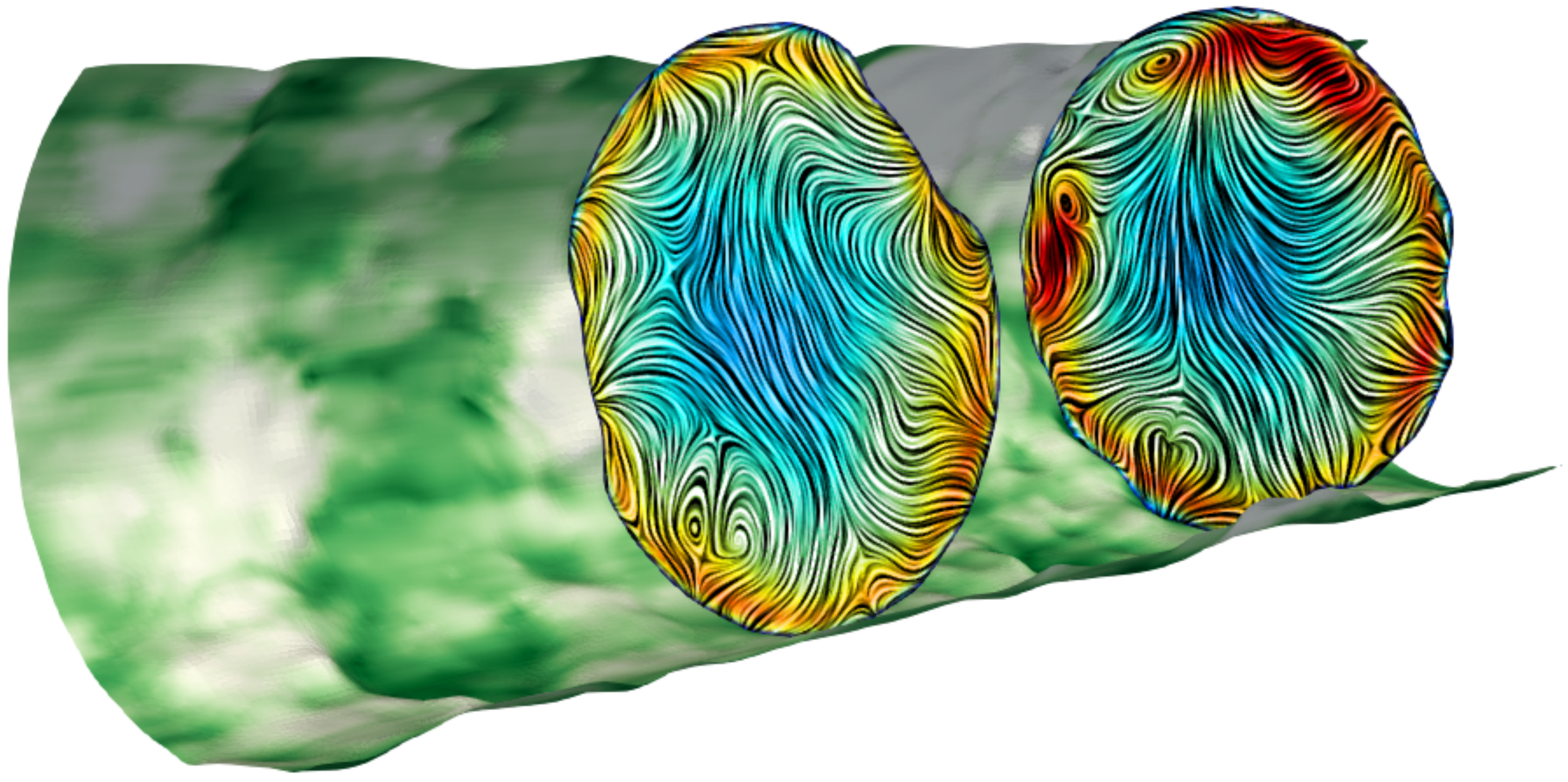}
         \label{fig:MeanLIC02d}}         
\subfloat[]{
\includegraphics[width=0.5\linewidth]{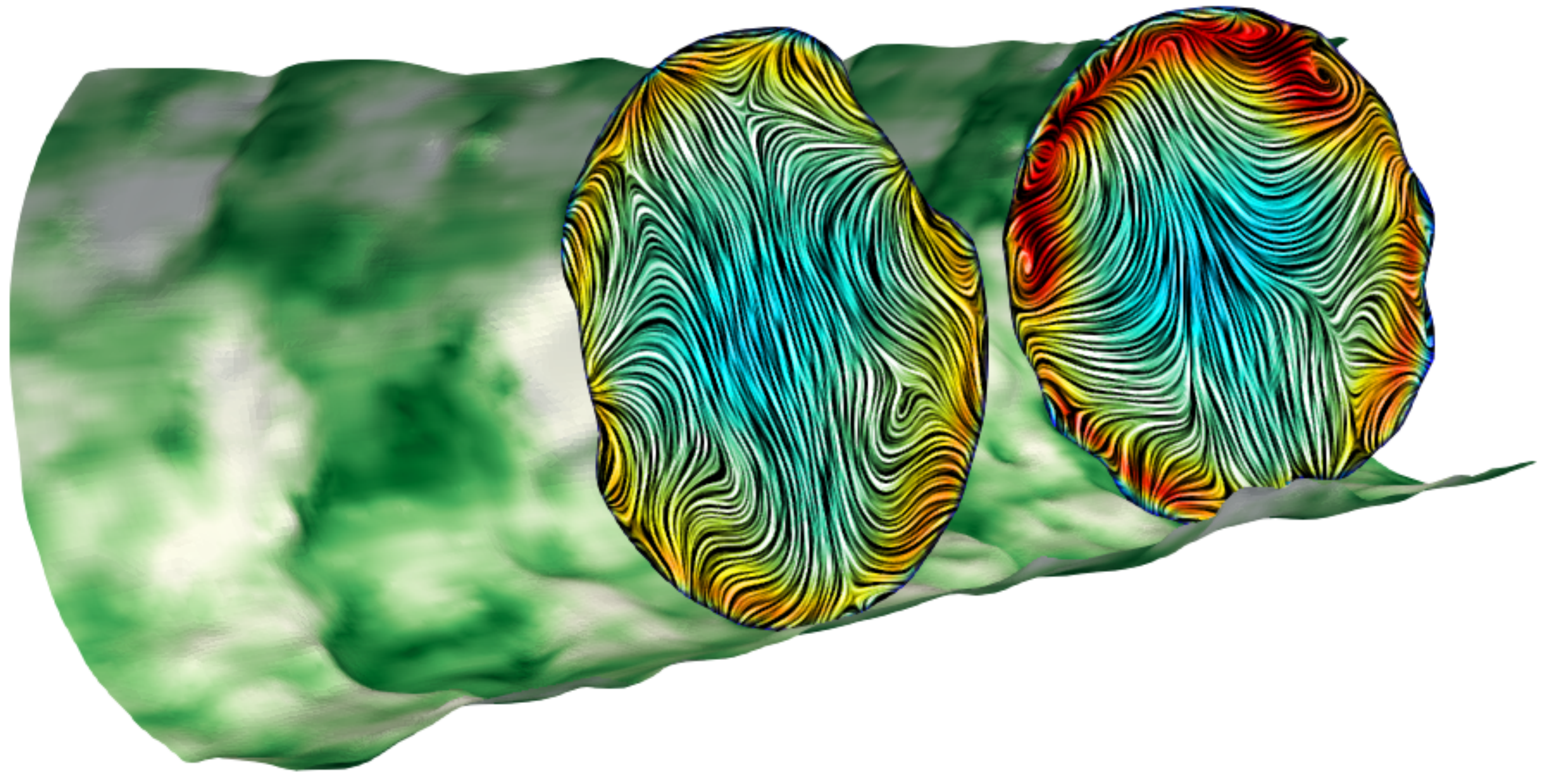}
         \label{fig:MeanLIC025d}}
         
\subfloat[]{
\includegraphics[width=0.5\linewidth]{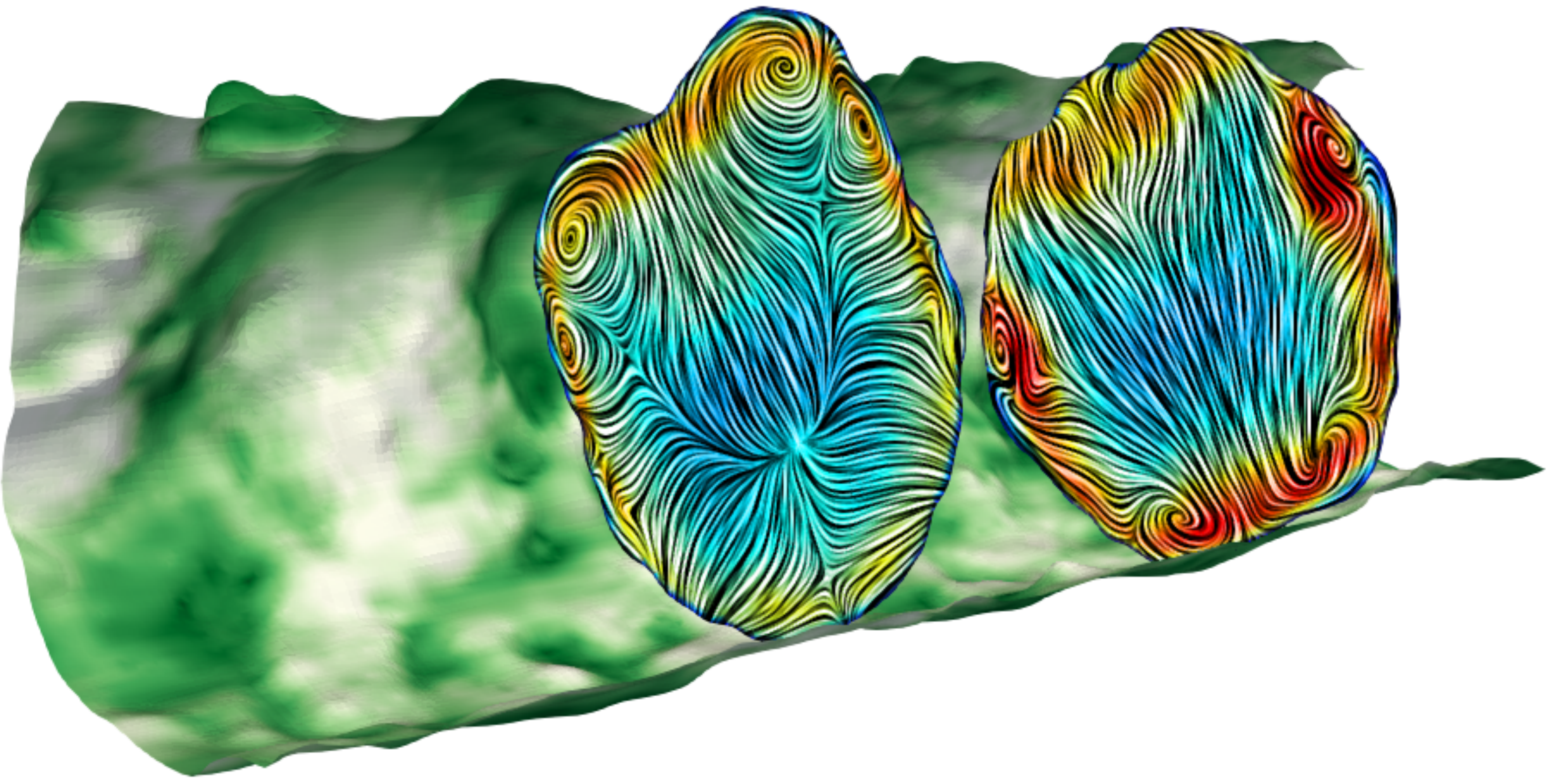}
         \label{fig:MeanLIC033d}}        
\subfloat[]{
\includegraphics[width=0.5\linewidth]{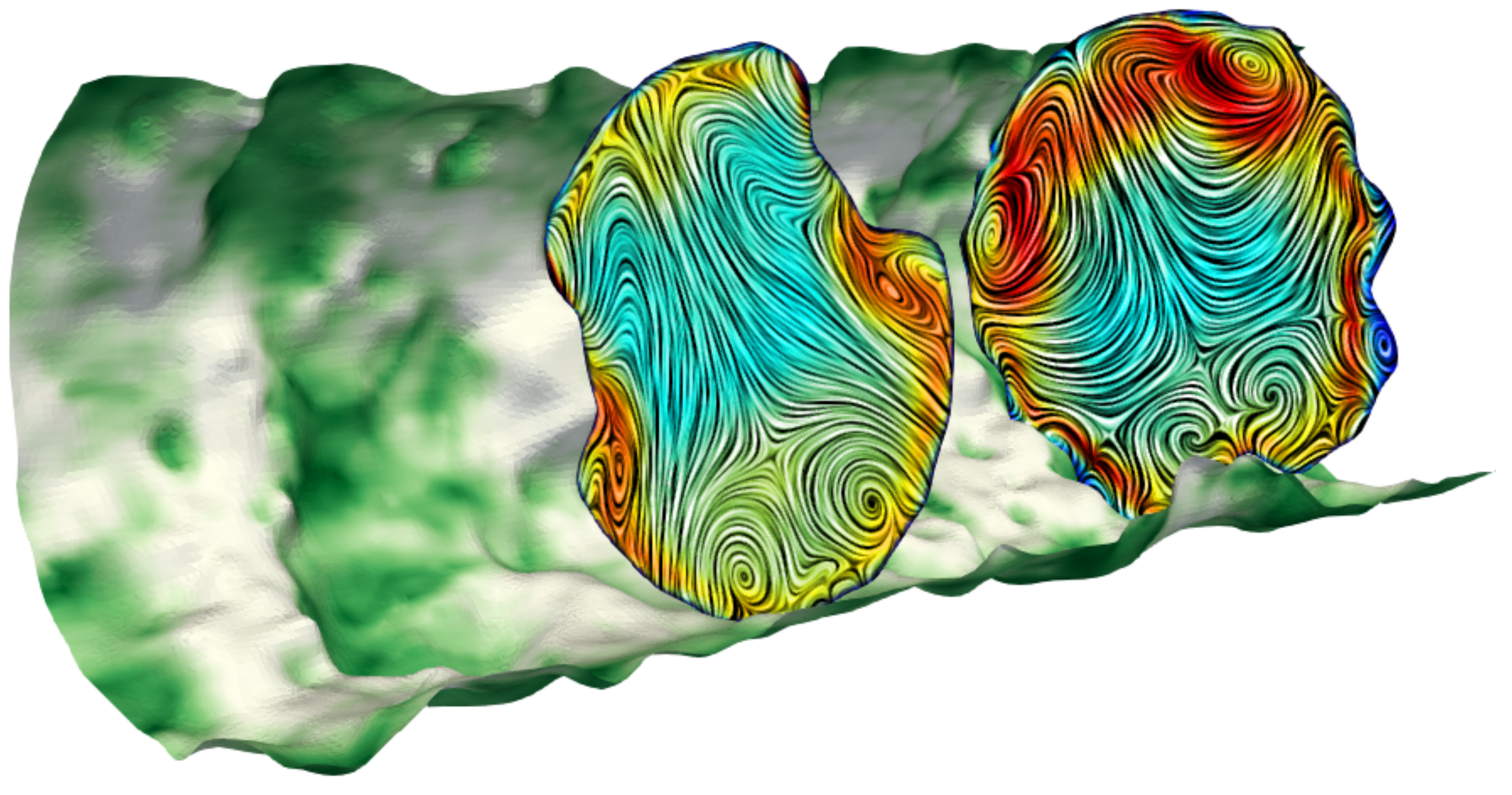}
         \label{fig:MeanLIC05d}}
\caption{Cross-sectional view of the root-mean-square of the streamwise velocity fluctuations field for different rough surfaces: \subref{fig:MeanLIC02d} C1, \subref{fig:MeanLIC025d} C2, \subref{fig:MeanLIC033d} C3, and \subref{fig:MeanLIC05d} C4. Streamlines obtained from the mean flow field show the development of the flow and secondary flow patterns attributed to wall roughness. In all plots, the wall is colored with the friction velocity, $u_\tau$.}
\label{fig: MeanSTLIC}
\end{center}
\end{figure*}

It is interesting to see how highly non-uniform and sparsely packed AM roughness induces the secondary flow motions and if there are some common features with regular or artificial roughness. We show the distribution of the instantaneous and root-mean-square (rms) velocity fluctuations overlayed with the secondary flow, depicted by in-plane streamlines for different rough surfaces, in Figs. \ref{fig: FlucSTLIC} (\textit{a})-(\textit{d}) and \ref{fig: MeanSTLIC} (\textit{a})-(\textit{d}), respectively. In the four cases of Fig. \ref{fig: FlucSTLIC} (\textit{a})-(\textit{d}), pronounced small-scale motions and counter-rotating pairs of vortices are encountered, resulting in momentum transport by ejecting and sweeping the fluid from and toward the wall. The size and frequency of these small-scale features are strictly dependent on the location of the plane and the time at which the snapshot is taken. From the instantaneous flow fields and the overlayed streamlines, it is very difficult to see if any secondary flow motion persists, and if so, then the impact of the roughness on the secondary motions is unclear. Therefore we look at the distribution of the streamwise rms velocity overlayed by the in-plane streamlines in Fig. \ref{fig: MeanSTLIC} (\textit{a})-(\textit{d}). Interestingly, out of four cases, pronounced secondary motion patterns can be observed mainly for C3 and C4 roughnesses. In the case of C3 roughness (Fig. \ref{fig: MeanSTLIC} (\textit{c})), the two main large-scale vortices originate from the recessed wall area. Two additional small vortices are located in the top and bottom regions of the wall, again close to the recessed area. On the other hand, in the case of C4 roughness (Fig. \ref{fig: MeanSTLIC} (\textit{d})), two main large-scale vortices originated from the recessed area of the wall, but were seldomly pushed away from the inner wall region. In addition, one can see four extra counter-rotating pairs of small structures on each side of the slightly bigger large-scale vortex. This is somewhat consistent with the results obtained by \citet{vanderwel2015,Hwang2018,wangsawijaya2020} for strip-type roughness and for ridge-type roughness, where the upwelling and downwelling movements were indicated above the elevated and recessed wall areas, respectively.

In contrast, in the case of C2 roughness (Fig. \ref{fig: MeanSTLIC} (\textit{b})), one can see the early development of some secondary flow structures, but for C1 roughness (Fig. \ref{fig: MeanSTLIC} (\textit{a})), only few inflection points are spotted. The origin of such inflexion points remains unanswered in this study and demands future investigations. Comparing these cases against the smooth surface pipe flow suggests that the roughness height, eventually surface topography, is an important parameter for secondary motion formation and destruction.
\subsection{Effect of surface anisotropy on the anisotropy of the Reynolds stress tensor}
\begin{figure*}
\begin{center}
\subfloat[Smooth Pipe]{
\includegraphics[width=0.42\linewidth]{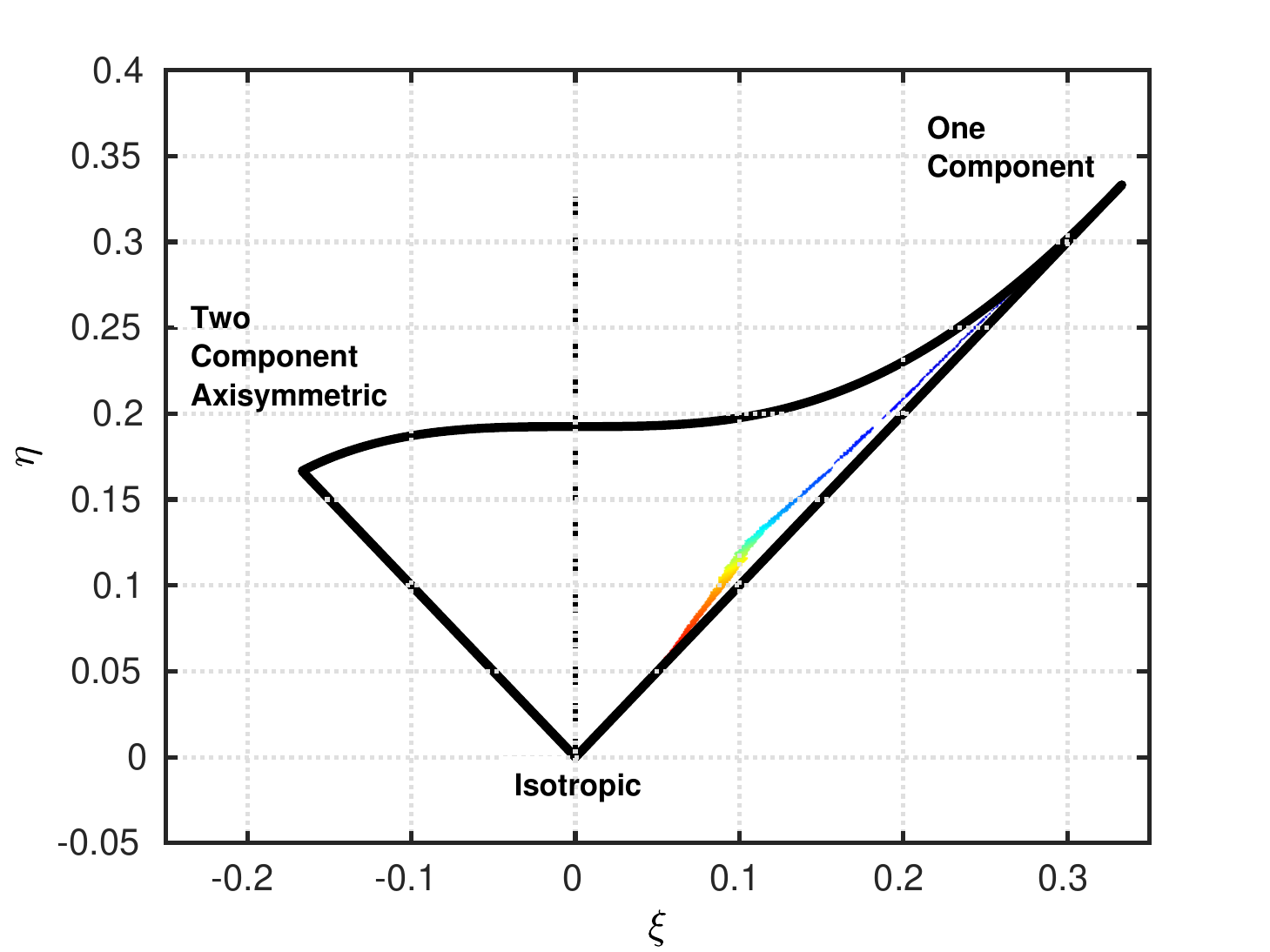}
}
\subfloat[C1]{
\includegraphics[width=0.42\linewidth]{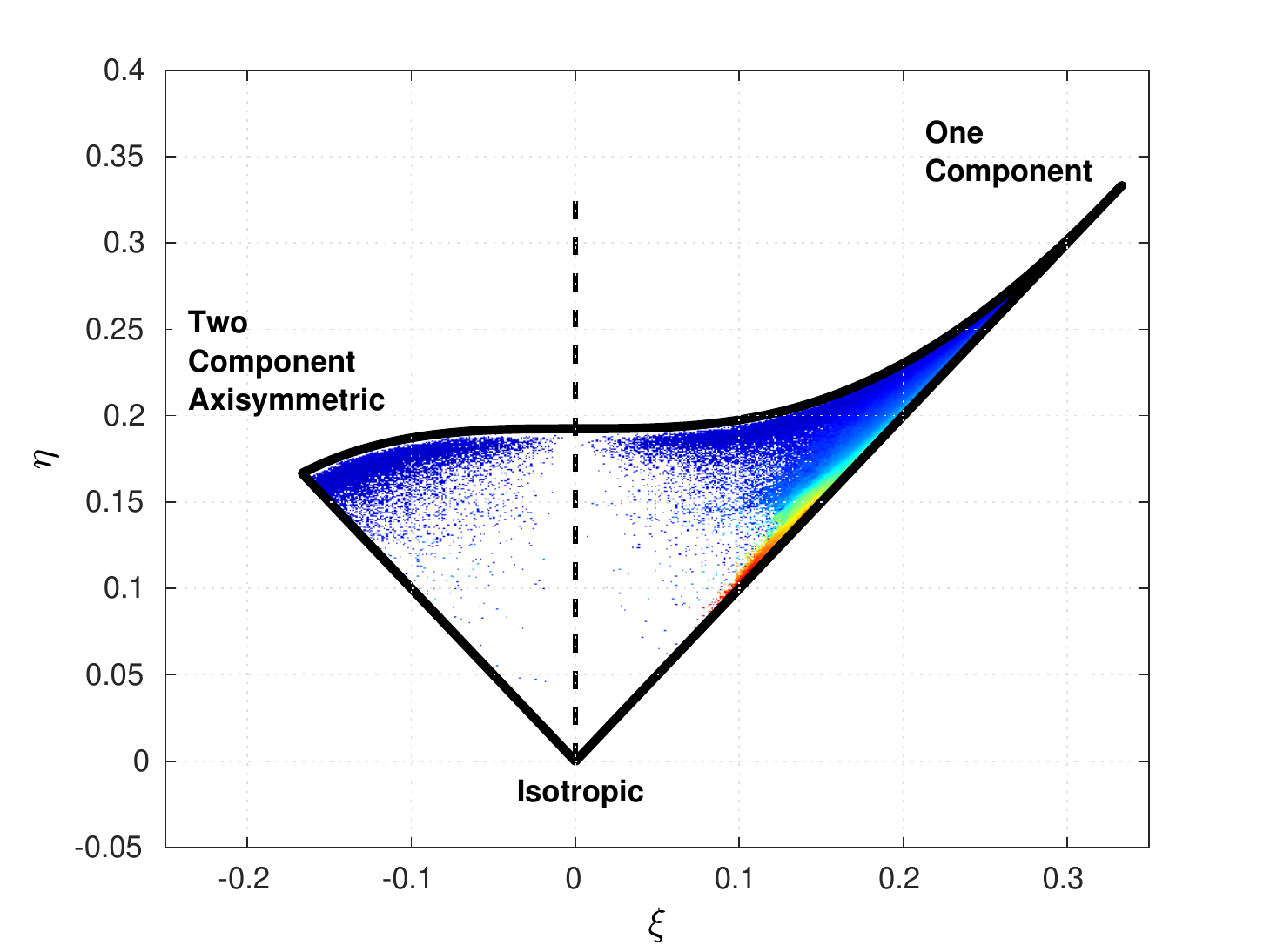}
}

\subfloat[C2]{
\includegraphics[width=0.42\linewidth]{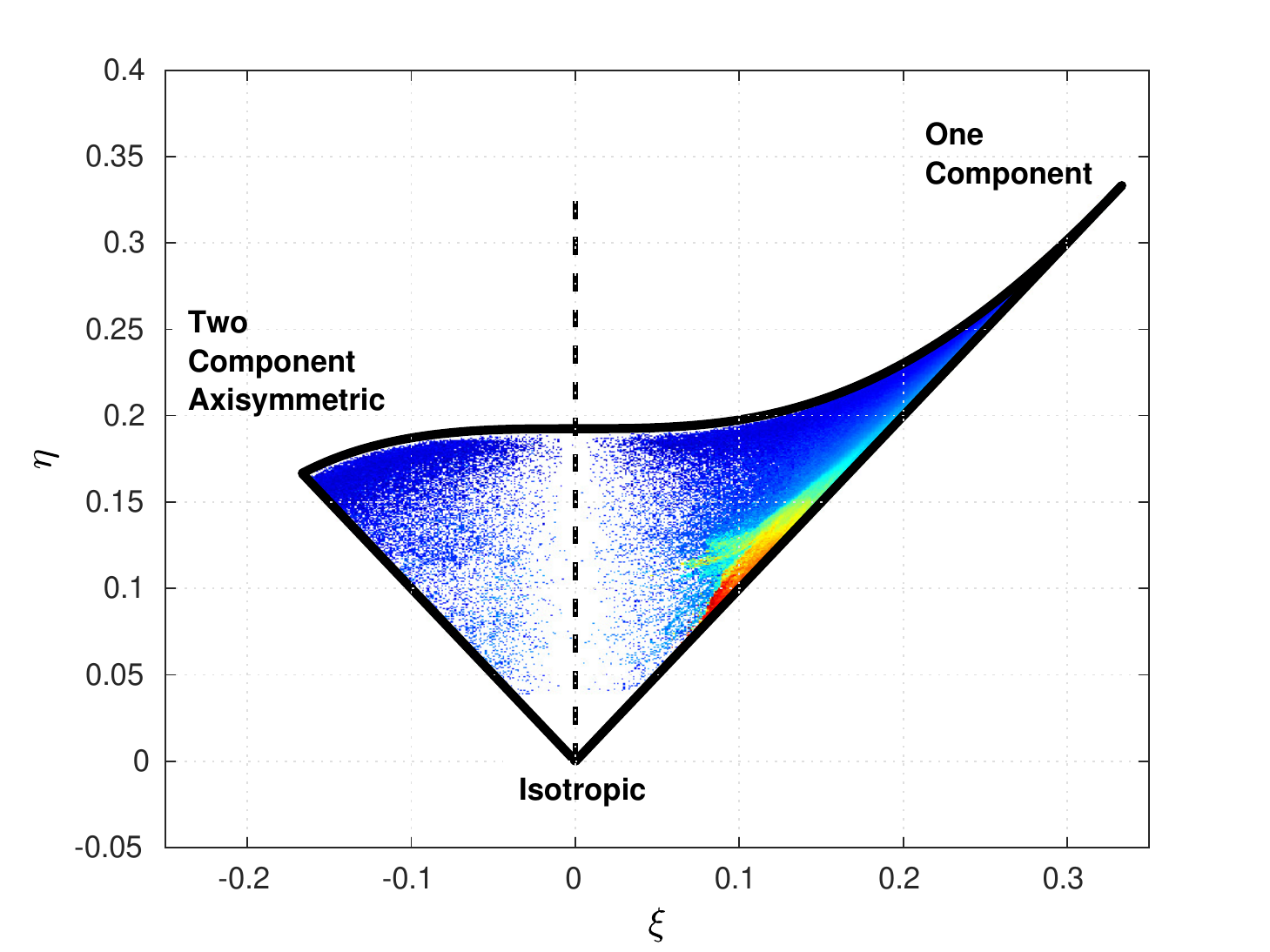}}
\subfloat[C3]{
\includegraphics[width=0.42\linewidth]{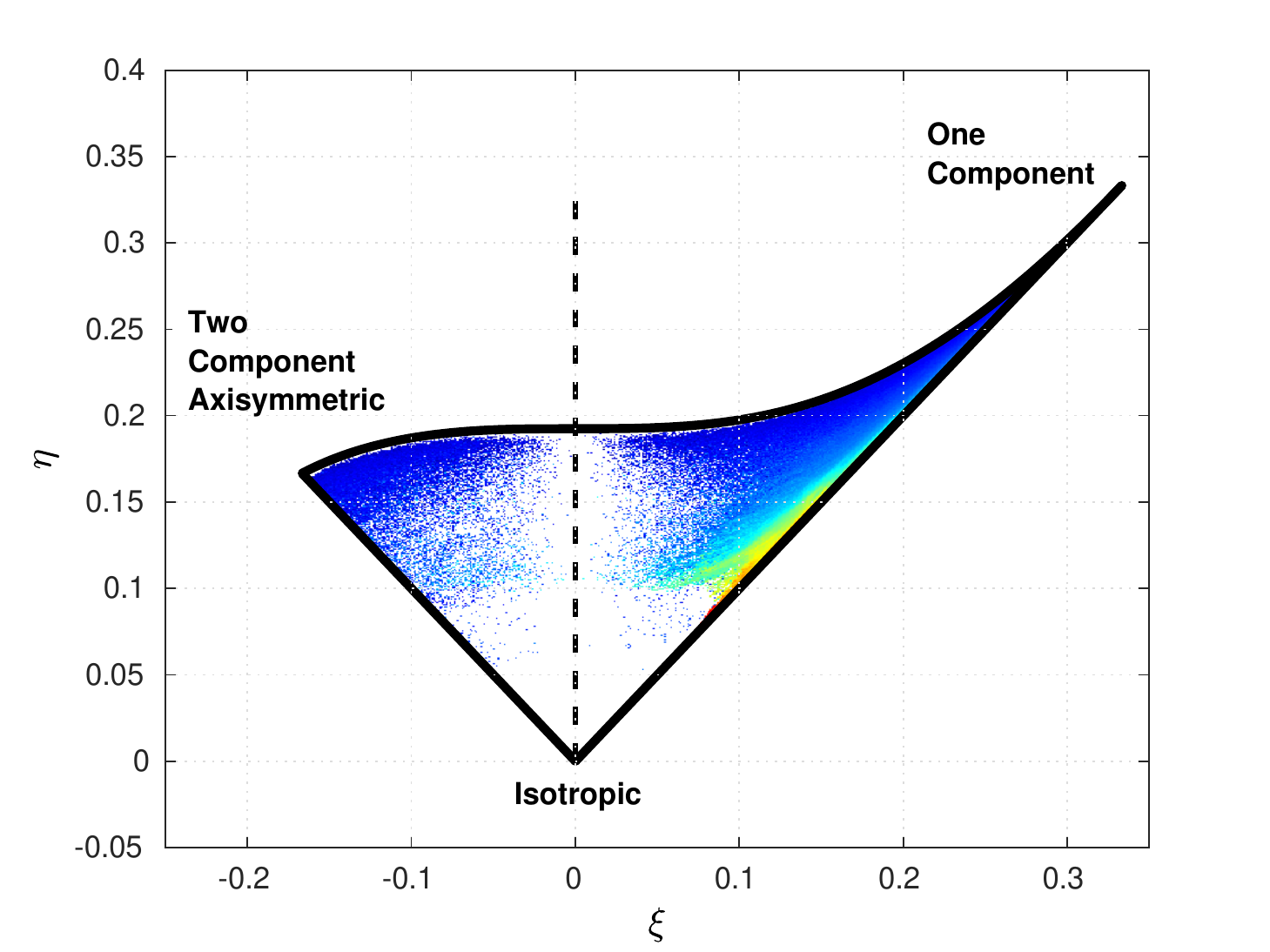}}

\subfloat[C4]{
\includegraphics[width=0.42\linewidth]{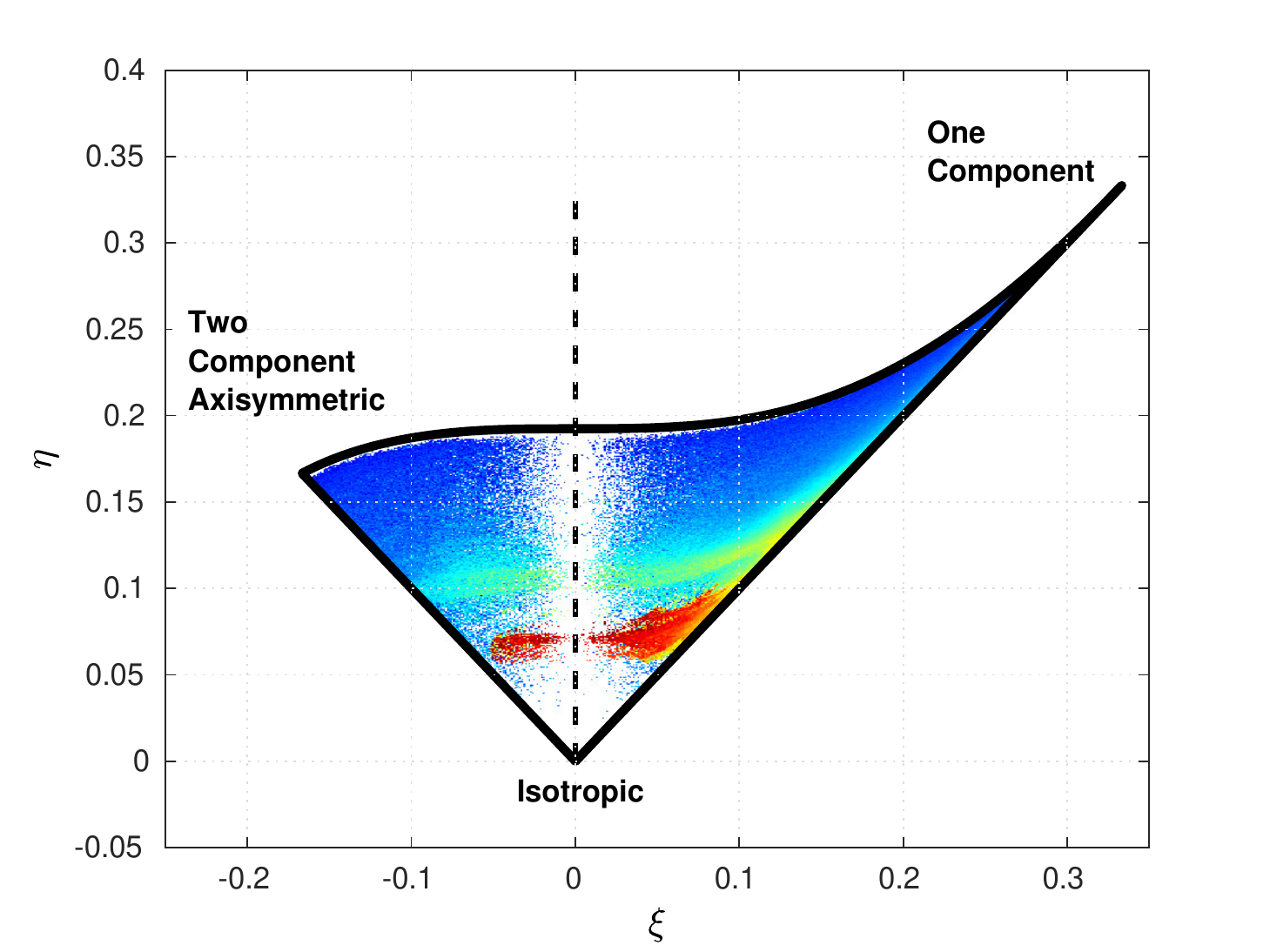}}
\caption{Anisotropy-invariant mapping of turbulence in turbulent rough pipe flow compiled from the present LES data at different roughness Reynolds numbers: (a) smooth pipe flow, $k_s^+ =0$, (b) $k_s^+=17.052$, (c) $k_s^+ =44.131$, (d) $k_s^+ =83.084$, and (e) $k_s^+ =422.45$. The data points for each case are based on all cells in the domain and colored with normalized wall distance values, $r^+$. Colormap varies from blue (minimum) to red (maximum). }
\label{fig: 5 ANISOTROPY MAPS FOR DIFFERENT SUBGRID MODELS}
\end{center}
\end{figure*}

The net anisotropy of the Reynolds stresses is commonly quantified using the second, $II_b$, and third, $III_b$, invariants of the normalized anisotropy tensor, $b_{ij}$, given by \cite{raupach1982averaging}
\begin{equation}
b_{ij} = \frac{\left< \overline{ u'_iu'_j}\right>}{\left< \overline{ u'_ku'_k}\right>} - \frac{1}{3} \delta_{ij}.
\label{anisotropy}
\end{equation}
The state of anisotropy can then be characterized with the two variables $\eta$ and $\zeta$ defined as
\begin{equation}
\eta^2 = -\frac{1}{3}II_b
\end{equation}
and
\begin{equation}
\zeta^3 = -\frac{1}{2}III_b.
\end{equation}
All realizable states of the Reynolds stress tensor are contained within a triangle in the $\left(\zeta,\eta\right)$ plane, the so-called Lumley triangle. All special turbulence cases can be characterized by the two invariants of $b_{ij}$ at the theoretical limits of the Lumley triangle. Details on the different realizable states can be found in Garg et al. \cite{Himani2022}. Lumley triangle plots for all rough surfaces are shown in Fig. \ref{fig: 5 ANISOTROPY MAPS FOR DIFFERENT SUBGRID MODELS}. We also show the Lumley triangle plot for smooth pipe flow as a reference. All cell-centered values of $\zeta$ and $\eta$ are post-processed using a probability density function (PDF) and represented with a colormap to match the wall-normal distance.  In the reference case of smooth-wall turbulent pipe flow, see Fig.  \ref{fig: 5 ANISOTROPY MAPS FOR DIFFERENT SUBGRID MODELS}(a) very close to the wall, the flow is close to a two-component state following the upper boundary of the Lumley triangle toward the one-component state denoted by the  upper right corner. The maximum anisotropy is reached for $r^+ \approx 8$. For $r^+>8$, the anisotropy decreases, and the $(\zeta, \eta)$ curve follows a path close to the right boundary of the Lumley triangle, indicating a close-to-axisymmetric, rod-like state and approaching, but not reaching, the state of maximum isotropy at the bottom summit of the Lumley triangle as $r^+$ increases. 

On the other hand, for the rough surfaces, the turbulent states are spread throughout the Lumley triangle, with the exception of the plain strain condition indicated by the dashed line, see Fig. \ref{fig: 5 ANISOTROPY MAPS FOR DIFFERENT SUBGRID MODELS}(b-e). In the deepest valleys of the surfaces, the Reynolds stress anisotropy tensor is close to a strongly anisotropic, one-component case. With increasing $r^+$, results follows a different path on the $(\zeta, \eta)$-map compared to the case for smooth-wall pipe flow. For all rough surfaces (C1-C4), the flow approaches the left side of the Lumley triangle, reaching an axisymmetric disk-like (ablate spheroid) state at the roughness mean plane. At this location, the streamwise and azimuthal Reynolds stresses are of comparable magnitude.  The anisotropy appears centered around the axisymmetric expansion and the two-component limit, indicating that one stress component is either larger than the other two or that the two components are similar in magnitude. An axisymmetric, disk-like state of the Reynolds stress anisotropy tensor is typical of mixing layers. Similar behavior has also been observed for turbulent flows over transverse bar roughness \cite{ashrafian2006structure}, k-type roughness \cite{smalley2002reynolds}, and irregular roughness \cite{busse2020influence}. Above the roughness mean plane, the path returns to the right side of the triangle, i.e., close to an axisymmetric, rod-like (prolate spheroid) state, and tracks the behavior of the smooth-wall case once the wall-normal coordinate exceeds the maximum roughness height. Finally, the most commonly occurring anisotropic states are axisymmetric expansion, one-component, two-component, and two-component axisymmetric states, and the probability of a certain turbulent state to occur increases with an increase in $k_s^+$.
\section{Conclusions}
We have presented a systematic numerical investigation of turbulent flows in Additively Manufactured (AM) rough surfaces employing roughness-resolved high-fidelity Large Eddy Simulations (LES) in OpenFOAM 7. The aim of the present simulations is twofold. First, shedding light on the effect of AM roughness on turbulent flow properties and highlighting the difference between existing regular and artificial roughnesses utilizing various statistical measures. Second, providing a rich database that will ease the development and assessment of rough-wall models.
The first challenging task in building this database is the generation of rough pipes from scanned planar AM rough surfaces and their characterization. To this end, we developed an open-source code to generate pipes from planar surfaces and to characterize the surface topographical properties.  
Four configurations of rough pipes from actual AM surfaces  with fixed skewness and kurtosis but different roughness height distributions and effective slopes have been constructed and simulated. 

By means of accurate spatial measurements, the numerical estimation of the roughness function is obtained for all cases and utilized to approximate equivalent sand-grain roughness height, $k_s$. The present AM rough surfaces yield a more significant value of roughness function compared to the literature results for regular and artificial roughnesses in a similar value range of $k_s$ \cite{shockling2006roughness,macdonald2019roughness,peeters2019turbulent}. The significant impact of $k_s$ on the skin friction coefficient is found through enhanced turbulence and drag forces.   
The Effective Slope, $ES$, which is directly linked to the alignment of the wall roughness with the flow, has a substantial impact on the flow topology, as expected. However, the present AM surfaces with $ES <0.35$ yield a significant value of roughness function, in contrast to regular and artificial roughnesses \cite{kuwata2020direct,saha2015scaling,de2016large,yuan2014estimation,napoli2008effect}. This could be a typical feature of AM surfaces and needs to be clarified in future investigations. In these cases, the existing empirical correlations find their limits, and new correlations are required. 
Furthermore, by defining an effective wall-normal distance, the collapse of the first- and second-order turbulence statistics is observed outside the roughness sublayer, validating Townsend's hypothesis for the present AM roughnesses. However, for one of the rough surfaces with $k_s/R_a=17.84$, the validity of Townsend's hypothesis is still questionable and needs future detailed investigations. Finally, a mixed boundary condition is proposed to improve axial fluctuation profiles' collapse in the roughness sublayer. 
We also attempted to understand, at least qualitatively, the impact of roughness topography on the generation of the secondary flow motions. Unfortunately, despite the abundance of results available for regular roughness, quantifying the AM roughness impact on secondary flow motion was not feasible in the current simulations. This calls for detailed future investigations of different types of roughnesses.
Lastly, we quantified the impact of roughness topography on turbulence anisotropy and compared it with smooth pipe flows. For the present rough surfaces, increasing $k_s$ results in high anisotropy. Furthermore, all turbulent states, i.e., one-component, two-component, and three-component states, existed with higher probabilities when increasing $k_s$, consistently with literature \cite{ashrafian2006structure,smalley2002reynolds,busse2020influence}.

We hope that the analysis reported in this work will help establish sound methodologies to compare in detail numerical simulations of AM rough surface flows to the available experimental and numerical results. This type of fine-scale comparison appears crucial to us in improving the
theoretical understanding of the subject and to validate the related numerical developments.

\section*{Acknowledgement}
The authors greatly appreciate the financial support provided by Vinnova under project number "2020-04 529" and Senior lecturer Martin Andersson (funding acquisition lead). Computer time was provided by the Swedish National Infrastructure for Computing (SNIC), partially funded by the Swedish Research Council through grant agreement no. 2018-05973. The authors also appreciate Siemens for providing the scanned AM surfaces and helpful discussions.
\section*{Author declarations}
\subsection*{Conflict of interest}
The authors have no conflicts of interest to disclose.
\subsection*{Data availability}
The data that support the findings of this study are available from the corresponding author upon reasonable request.
%
\end{document}